\documentclass[11pt]{article}

\usepackage{amsmath}
\usepackage{amssymb}
\usepackage{graphicx}
\usepackage{program}
\usepackage{ifthen}
\usepackage{array}
\usepackage{color}
\usepackage[table]{xcolor}
\usepackage{hyperref}
\usepackage{xspace}
\usepackage{listings}
\usepackage{subcaption}

\definecolor{darkgreen}{rgb}{0,0.44,0}
\definecolor{darkred}{rgb}{0.44,0,0}
\definecolor{darkblue}{rgb}{0,0,0.44}

\definecolor{myred}{rgb}{0.82,0.0,0.0} 
\definecolor{mygreen}{rgb}{0.0,0.44,0.0} 
\definecolor{myblue}{rgb}{0.0,0.0,0.82} 
\definecolor{mygray}{rgb}{0.4,0.4,0.4}
\definecolor{mymauve}{rgb}{0.58,0,0.82}
\definecolor{mycreme}{rgb}{1.0,0.8,0.2} 

\newcommand{\crr}[1]{\textcolor{myred}{#1}}
\newcommand{\crb}[1]{\textcolor{myblue}{#1}}
\newcommand{\crg}[1]{\textcolor{mygreen}{#1}}
\newcommand{\crc}[1]{\textcolor{mycreme}{#1}}

\usepackage{arydshln}

\newcolumntype{I}{!{\vrule width 1.5pt}}

\newlength\savedwidth 

\newcommand\whline{\noalign{\global\savedwidth\arrayrulewidth 
                            \global\arrayrulewidth 1.5pt}%
           \hline 
           \noalign{\global\arrayrulewidth\savedwidth}}

\newcommand{\routinename}{}

\newcommand{\precondition}{~}

\newcommand{\postcondition}{~}

\newcommand{\invariant}{~}

\newcommand{\guard}{~}

\newcommand{\partitionings}{~}

\newcommand{\partitionsizes}{~}

\newcommand{\blocksize}{blank}

\newcommand{\repartitionings}{~}

\newcommand{\repartitionsizes}{~}

\newcommand{\moveboundaries}{~}

\newcommand{\beforeupdate}{~}

\newcommand{\afterupdate}{~}

\newcommand{\update}{~}

\newcommand{\NoShow}[1]{}

\newcommand{\FlaAlgorithm}{
\begin{tabular}{| p{0.92\textwidth}|} \hline
$\mbox{\color{blue}Algorithm:~}\routinename$
\\ \whline
\partitionings \\
$\mbox{\color{blue} ~~~where~}$ \partitionsizes 
\\ 
$\mbox{\color{blue}while~} \guard \mbox{~\color{blue} do}$
\\
\ifthenelse{\equal{\blocksize}{1}}{\\}%
{%
\ifthenelse{ \equal{\blocksize}{blank} }{}%
{~~~~{\bf Determine block size $ \blocksize $}\\}%
}
~~~~ 
\repartitionings \\
~~~$\mbox{\color{blue} ~~~where~}$ \repartitionsizes
\\ \hline
~~~~  \update 
\\ \hline
~~~~ 
\moveboundaries 
\\
$\mbox{\color{blue} endwhile} $
\\ \hline 
\end{tabular}
}

\newcommand{\FlaWorksheet}{
\begin{tabular}{| c | p{0.98\textwidth} |}\hline
Step & $\mbox{\color{blue}Algorithm:~}\routinename$
\\ \hline
\rowcolor{yellow!75}
1a & $ \precondition $ 
\\ \whline
4 & 
\begin{minipage}[t]{0.9\textwidth}%
\partitionings~ \\
$\mbox{\color{blue} ~~~where~}$ \partitionsizes
\end{minipage}
\\ \hline
\rowcolor{yellow!75}
2 & $ \invariant $ 
\\ \hline
3 &$\mbox{\color{blue}while~} \guard \mbox{~\color{blue} do}$
\\ \hline 
\rowcolor{yellow!75}
2,3 & ~~~~ $ \invariant \wedge \guard$ 
\\ \hline
5a & ~~~~ \begin{minipage}[t]{0.85\textwidth}%
\ifthenelse{\equal{\blocksize}{1}}{}%
{%
\ifthenelse{ \equal{\blocksize}{blank} }{}%
{{\bf Determine block size $ \blocksize $}\\}%
}
\repartitionings~ \\
$\mbox{\color{blue} ~~~where~}$ \repartitionsizes
\end{minipage}
\\ \hline
\rowcolor{yellow!75}
6 & ~~~~ $\beforeupdate $
\\ \hline
8 & ~~~~  \update 
\\ \hline
5b & ~~~~ \begin{minipage}[t]{0.85\textwidth}%
\moveboundaries~
\end{minipage}
\\ \hline
\rowcolor{yellow!75}
7 & ~~~~ $\afterupdate $
\\ \hline
\rowcolor{yellow!75}
2 & ~~~~ $ \invariant  $ 
\\ \hline
 &$\mbox{\color{blue} endwhile} $
\\ \hline \whline
\rowcolor{yellow!75}
2,3 & $ \invariant \wedge \neg( \guard )$ 
\\ \hline
\rowcolor{yellow!75}
1b & $ \postcondition $ 
\\ \hline
\end{tabular}
}

\newcommand{\FlaWorksheetNine}{
\begin{tabular}{| c | p{0.98\textwidth} |}\hline
Step & $\mbox{\color{blue}Algorithm:~}\routinename$
\\ \hline
\rowcolor{yellow!75}
\phantom{1a} & $ \phantom\precondition $ 
\\ \whline
\phantom{4} & 
\begin{minipage}[t]{0.9\textwidth}%
\partitionings~ \\
$\mbox{\color{blue} ~~~where~}$ \partitionsizes
\end{minipage}
\\ \hline
\rowcolor{yellow!75}
\phantom{2} & $ \phantom\invariant $ 
\\ \hline
\phantom{3} &$\mbox{\color{blue}while~} \guard \mbox{~\color{blue} do}$
\\ \hline 
\rowcolor{yellow!75}
\phantom{2,3} & ~~~~ $ \phantom\invariant \phantom \wedge \phantom
                \guard $ 
\\ \hline
\phantom{5a} & ~~~~ \begin{minipage}[t]{0.85\textwidth}%
\ifthenelse{\equal{\blocksize}{1}}{}%
{%
\ifthenelse{ \equal{\blocksize}{blank} }{}%
{{\bf Determine block size $ \blocksize $}\\}%
}
\repartitionings~ \\
$\mbox{\color{blue} ~~~where~}$ \repartitionsizes
\end{minipage}
\\ \hline
\rowcolor{yellow!75}
\phantom{6} & ~~~~ $\phantom\beforeupdate $
\\ \hline
\phantom{8} & ~~~~  \update 
\\ \hline
\phantom{5b} & ~~~~ \begin{minipage}[t]{0.85\textwidth}%
\moveboundaries~
\end{minipage}
\\ \hline
\rowcolor{yellow!75}
\phantom{7} & ~~~~ $\phantom\afterupdate $
\\ \hline
\rowcolor{yellow!75}
\phantom{2} & ~~~~ $ \phantom\invariant  $ 
\\ \hline
 &$\mbox{\color{blue} endwhile} $
\\ \hline \whline
\rowcolor{yellow!75}
\phantom{2,3} & $ \phantom\invariant \wedge \neg( \phantom\guard )$ 
\\ \hline
\rowcolor{yellow!75}
\phantom{1b} & $ \phantom\postcondition $ 
\\ \hline
\end{tabular}
}

\newcommand{\FlaWorksheetEight}{
\begin{tabular}{| c | p{0.98\textwidth} |}\hline
Step & $\mbox{\color{blue}Algorithm:~}\routinename$
\\ \hline
\rowcolor{yellow!75}
1a & $ \precondition $ 
\\ \whline
4 & 
\begin{minipage}[t]{0.9\textwidth}%
\partitionings~ \\
$\mbox{\color{blue} ~~~where~}$ \partitionsizes
\end{minipage}
\\ \hline
\rowcolor{yellow!75}
2 & $ \invariant $ 
\\ \hline
3 &$\mbox{\color{blue}while~} \guard \mbox{~\color{blue} do}$
\\ \hline 
\rowcolor{yellow!75}
2,3 & ~~~~ $ \invariant \wedge \guard $ 
\\ \hline
5a & ~~~~ \begin{minipage}[t]{0.85\textwidth}%
\ifthenelse{\equal{\blocksize}{1}}{}%
{%
\ifthenelse{ \equal{\blocksize}{blank} }{}%
{{\bf Determine block size $ \blocksize $}\\}%
}
\repartitionings~ \\
$\mbox{\color{blue} ~~~where~}$ \repartitionsizes
\end{minipage}
\\ \hline
\rowcolor{yellow!75}
6 & ~~~~ $\beforeupdate $
\\ \hline
\rowcolor{orange!50}    
8 & ~~~~  \update 
\\ \hline
5b & ~~~~ \begin{minipage}[t]{0.85\textwidth}%
\moveboundaries~
\end{minipage}
\\ \hline
\rowcolor{yellow!75}
7 & ~~~~ $\afterupdate $
\\ \hline
\rowcolor{yellow!75}
2 & ~~~~ $ \invariant  $ 
\\ \hline
 &$\mbox{\color{blue} endwhile} $
\\ \hline \whline
\rowcolor{yellow!75}
2,3 & $ \invariant \wedge \neg( \guard )$ 
\\ \hline
\rowcolor{yellow!75}
1b & $ \postcondition $ 
\\ \hline
\end{tabular}
}

\newcommand{\FlaWorksheetSeven}{
\begin{tabular}{| c | p{0.98\textwidth} |}\hline
Step & $\mbox{\color{blue}Algorithm:~}\routinename$
\\ \hline
\rowcolor{yellow!75}
1a & $ \precondition $ 
\\ \whline
4 & 
\begin{minipage}[t]{0.9\textwidth}%
\partitionings~ \\
$\mbox{\color{blue} ~~~where~}$ \partitionsizes
\end{minipage}
\\ \hline
\rowcolor{yellow!75}
2 & $ \invariant $ 
\\ \hline
3 &$\mbox{\color{blue}while~} \guard \mbox{~\color{blue} do}$
\\ \hline 
\rowcolor{yellow!75}
2,3 & ~~~~ $ \invariant \wedge \guard$ 
\\ \hline
5a & ~~~~ \begin{minipage}[t]{0.85\textwidth}%
\ifthenelse{\equal{\blocksize}{1}}{}%
{%
\ifthenelse{ \equal{\blocksize}{blank} }{}%
{{\bf Determine block size $ \blocksize $}\\}%
}
\repartitionings~ \\
$\mbox{\color{blue} ~~~where~}$ \repartitionsizes
\end{minipage}
\\ \hline
\rowcolor{yellow!75}
6 & ~~~~ $\beforeupdate $
\\ \hline
8 & ~~~~  \phantom\update 
\\ \hline
5b & ~~~~ \begin{minipage}[t]{0.85\textwidth}%
\moveboundaries~
\end{minipage}
\\ \hline
\rowcolor{orange!50}    
7 & ~~~~ $\afterupdate $
\\ \hline
\rowcolor{yellow!75}
2 & ~~~~ $ \invariant  $ 
\\ \hline
 &$\mbox{\color{blue} endwhile} $
\\ \hline \whline
\rowcolor{yellow!75}
2 & $ \invariant \wedge \neg( \guard )$ 
\\ \hline
\rowcolor{yellow!75}
1b & $ \postcondition $ 
\\ \hline
\end{tabular}
}

\newcommand{\FlaWorksheetSix}{
\begin{tabular}{| c | p{0.98\textwidth} |}\hline
Step & $\mbox{\color{blue}Algorithm:~}\routinename$
\\ \hline
\rowcolor{yellow!75}
1a & $ \precondition $ 
\\ \whline
4 & 
\begin{minipage}[t]{0.9\textwidth}%
\partitionings~ \\
$\mbox{\color{blue} ~~~where~}$ \partitionsizes
\end{minipage}
\\ \hline
\rowcolor{yellow!75}
2 & $ \invariant $ 
\\ \hline
3 &$\mbox{\color{blue}while~} \guard \mbox{~\color{blue} do}$
\\ \hline 
\rowcolor{yellow!75}
2,3 & ~~~~ $ \invariant \wedge \guard $ 
\\ \hline
5a & ~~~~ \begin{minipage}[t]{0.85\textwidth}%
\ifthenelse{\equal{\blocksize}{1}}{}%
{%
\ifthenelse{ \equal{\blocksize}{blank} }{}%
{{\bf Determine block size $ \blocksize $}\\}%
}
\repartitionings~ \\
$\mbox{\color{blue} ~~~where~}$ \repartitionsizes
\end{minipage}
\\ \hline
\rowcolor{orange!50}   
6 & ~~~~ $\beforeupdate $
\\ \hline
8 & ~~~~  \phantom\update 
\\ \hline
5b & ~~~~ \begin{minipage}[t]{0.85\textwidth}%
\moveboundaries~
\end{minipage}
\\ \hline
\rowcolor{yellow!75}
7 & ~~~~ $\phantom\afterupdate $
\\ \hline
\rowcolor{yellow!75}
2 & ~~~~ $ \invariant  $ 
\\ \hline
 &$\mbox{\color{blue} endwhile} $
\\ \hline \whline
\rowcolor{yellow!75}
2,3 & $ \invariant \wedge \neg( \guard )$ 
\\ \hline
\rowcolor{yellow!75}
1b & $ \postcondition $ 
\\ \hline
\end{tabular}
}

\newcommand{\FlaWorksheetFive}{
\begin{tabular}{| c | p{0.98\textwidth} |}\hline
Step & $\mbox{\color{blue}Algorithm:~}\routinename$
\\ \hline
\rowcolor{yellow!75}
1a & $ \precondition $ 
\\ \whline
4 & 
\begin{minipage}[t]{0.9\textwidth}%
\partitionings~ \\
$\mbox{\color{blue} ~~~where~}$ \partitionsizes
\end{minipage}
\\ \hline
\rowcolor{yellow!75}
2 & $ \invariant $ 
\\ \hline
3 &$\mbox{\color{blue}while~} \guard \mbox{~\color{blue} do}$
\\ \hline 
\rowcolor{yellow!75}
2,3 & ~~~~ $ \invariant \wedge \guard $ 
\\ \hline
\rowcolor{orange!50}   
5a & ~~~~ \begin{minipage}[t]{0.85\textwidth}%
\ifthenelse{\equal{\blocksize}{1}}{}%
{%
\ifthenelse{ \equal{\blocksize}{blank} }{}%
{{\bf Determine block size $ \blocksize $}\\}%
}
\repartitionings~ \\
$\mbox{\color{blue} ~~~where~}$ \repartitionsizes
\end{minipage}
\\ \hline
\rowcolor{yellow!75}
6 & ~~~~ $\phantom\beforeupdate $
\\ \hline
8 & ~~~~  \phantom\update 
\\ \hline
\rowcolor{orange!50}   
5b & ~~~~ \begin{minipage}[t]{0.85\textwidth}%
\moveboundaries~
\end{minipage}
\\ \hline
\rowcolor{yellow!75}
7 & ~~~~ $\phantom\afterupdate $
\\ \hline
\rowcolor{yellow!75}
2 & ~~~~ $ \invariant  $ 
\\ \hline
 &$\mbox{\color{blue} endwhile} $
\\ \hline \whline
\rowcolor{yellow!75}
2,3 & $ \invariant \wedge \neg( \guard )$ 
\\ \hline
\rowcolor{yellow!75}
1b & $ \postcondition $ 
\\ \hline
\end{tabular}
}

\newcommand{\FlaWorksheetFour}{
\begin{tabular}{| c | p{0.98\textwidth} |}\hline
Step & $\mbox{\color{blue}Algorithm:~}\routinename$
\\ \hline
\rowcolor{yellow!75}
1a & $ \precondition $ 
\\ \whline
\rowcolor{orange!50}   
4 & 
\begin{minipage}[t]{0.9\textwidth}%
\partitionings~ \\
$\mbox{\color{blue} ~~~where~}$ \partitionsizes
\end{minipage}
\\ \hline
\rowcolor{yellow!75}
2 & $ \invariant $ 
\\ \hline
3 &$\mbox{\color{blue}while~} \guard \mbox{~\color{blue} do}$
\\ \hline 
\rowcolor{yellow!75}
2,3 & ~~~~ $ \invariant \wedge \guard $ 
\\ \hline
5a & ~~~~ \begin{minipage}[t]{0.85\textwidth}%
\ifthenelse{\equal{\blocksize}{1}}{}%
{%
\ifthenelse{ \equal{\blocksize}{blank} }{}%
{{\bf Determine block size $ \phantom\blocksize $}\\}%
}
$\mbox{\phantom\repartitionings}$~ \\
$\mbox{\color{blue} ~~~where~}$ \phantom\repartitionsizes
\end{minipage}
\\ \hline
\rowcolor{yellow!75}
6 & ~~~~ $\phantom\beforeupdate $
\\ \hline
8 & ~~~~  \phantom\update 
\\ \hline
5b & ~~~~ \begin{minipage}[t]{0.85\textwidth}%
\phantom\moveboundaries~
\end{minipage}
\\ \hline
\rowcolor{yellow!75}
7 & ~~~~ $\phantom\afterupdate $
\\ \hline
\rowcolor{yellow!75}
2 & ~~~~ $ \invariant  $ 
\\ \hline
 &$\mbox{\color{blue} endwhile} $
\\ \hline \whline
\rowcolor{yellow!75}
2,3 & $ \invariant \wedge \neg( \guard )$ 
\\ \hline
\rowcolor{yellow!75}
1b & $ \postcondition $ 
\\ \hline
\end{tabular}
}

\newcommand{\FlaWorksheetThree}{
\begin{tabular}{| c | p{0.98\textwidth} |}\hline
Step & $\mbox{\color{blue}Algorithm:~}\routinename$
\\ \hline
\rowcolor{yellow!75}
1a & $ \precondition $ 
\\ \whline
4 & 
\begin{minipage}[t]{0.9\textwidth}%
$\mbox{\phantom{\partitionings}}$~ \\
$\mbox{\color{blue} ~~~where~}$\phantom{\partitionsizes}  
\end{minipage}
\\ \hline
\rowcolor{yellow!75}
2 & $ \invariant $ 
\\ \hline
\rowcolor{orange!50}  
3 &$\mbox{\color{blue}while~} \guard \mbox{~\color{blue} do}$
\\ \hline 
\rowcolor{orange!50}   
2,3 & ~~~~ $ \invariant \wedge \guard $ 
\\ \hline
5a & ~~~~ \begin{minipage}[t]{0.85\textwidth}%
\ifthenelse{\equal{\blocksize}{1}}{}%
{%
\ifthenelse{ \equal{\blocksize}{blank} }{}%
{{\bf Determine block size $ \phantom\blocksize $}\\}%
}
$\mbox{\phantom\repartitionings}$~ \\
$\mbox{\color{blue} ~~~where~}$ \phantom\repartitionsizes
\end{minipage}
\\ \hline
\rowcolor{yellow!75}
6 & ~~~~ $\phantom\beforeupdate $
\\ \hline
8 & ~~~~  \phantom\update 
\\ \hline
5b & ~~~~ \begin{minipage}[t]{0.85\textwidth}%
\phantom\moveboundaries~
\end{minipage}
\\ \hline
\rowcolor{yellow!75}
7 & ~~~~ $\phantom\afterupdate $
\\ \hline
\rowcolor{yellow!75}
2 & ~~~~ $ \invariant  $ 
\\ \hline
 &$\mbox{\color{blue} endwhile} $
\\ \hline \whline
\rowcolor{orange!50}   
2,3 & $ \invariant \wedge \neg( \guard )$ 
\\ \hline
\rowcolor{yellow!75}
1b & $ \postcondition $ 
\\ \hline
\end{tabular}
}

\newcommand{\FlaWorksheetTwo}{
\begin{tabular}{| c | p{0.98\textwidth} |}\hline
Step & $\mbox{\color{blue}Algorithm:~}\routinename$
\\ \hline
\rowcolor{yellow!75}
1a & $ \precondition $ 
\\ \whline
4 & 
\begin{minipage}[t]{0.9\textwidth}%
$\mbox{\phantom{\partitionings}}$~ \\
$\mbox{\color{blue} ~~~where~} $ \phantom{\partitionsizes} 
\end{minipage}
\\ \hline
\rowcolor{orange!50} 
2 & $ \invariant $ 
\\ \hline
3 &$\mbox{\color{blue}while~} \phantom\guard \mbox{~\color{blue} do}$
\\ \hline 
\rowcolor{orange!50} 
2,3 & ~~~~ $ \invariant \wedge \phantom \guard $ 
\\ \hline
5a & ~~~~ \begin{minipage}[t]{0.85\textwidth}%
\ifthenelse{\equal{\blocksize}{1}}{}%
{%
\ifthenelse{ \equal{\blocksize}{blank} }{}%
{{\bf Determine block size $ \phantom\blocksize $}\\}%
}
$\mbox{\phantom\repartitionings}$~ \\
$\mbox{\color{blue} ~~~where~}$ \phantom\repartitionsizes
\end{minipage}
\\ \hline
\rowcolor{yellow!75}
6 & ~~~~ $\phantom\beforeupdate $
\\ \hline
8 & ~~~~  \phantom\update 
\\ \hline
5b & ~~~~ \begin{minipage}[t]{0.85\textwidth}%
\phantom\moveboundaries~
\end{minipage}
\\ \hline
\rowcolor{yellow!75}
7 & ~~~~ $\phantom\afterupdate $
\\ \hline
\rowcolor{orange!50} 
2 & ~~~~ $ \invariant  $ 
\\ \hline
 &$\mbox{\color{blue} endwhile} $
\\ \hline \whline
\rowcolor{orange!50} 
2 & $ \invariant \wedge \neg( \phantom\guard )$ 
\\ \hline
\rowcolor{yellow!75}
1b & $ \postcondition $ 
\\ \hline
\end{tabular}
}

\newcommand{\FlaWorksheetOne}{
\begin{tabular}{| c | p{0.98\textwidth} |}\hline
Step & $\mbox{\color{blue}Algorithm:~}\routinename$
\\ \hline
\rowcolor{orange!50}
1a & $ \precondition $ 
\\ \whline
4 & 
\begin{minipage}[t]{0.9\textwidth}%
$\mbox{\phantom{\partitionings}}$ ~ \\
$\mbox{\color{blue} ~~~where~}$ \phantom{\partitionsizes} 
\end{minipage}
\\ \hline
\rowcolor{yellow!75}
2 & $ \phantom\invariant $ 
\\ \hline
3 &$\mbox{\color{blue}while~} \phantom\guard \mbox{~\color{blue} do}$
\\ \hline 
\rowcolor{yellow!75}
2,3 & ~~~~ $ \phantom\invariant \wedge \phantom \guard$ 
\\ \hline
5a & ~~~~ \begin{minipage}[t]{0.85\textwidth}%
\ifthenelse{\equal{\blocksize}{1}}{}%
{%
\ifthenelse{ \equal{\blocksize}{blank} }{}%
{{\bf Determine block size $ \phantom\blocksize $}\\}%
}
$\mbox{\phantom\repartitionings}$ ~ \\
$\mbox{\color{blue} ~~~where~}$ \phantom\repartitionsizes
\end{minipage}
\\ \hline
\rowcolor{yellow!75}
6 & ~~~~ $\phantom\beforeupdate $
\\ \hline
8 & ~~~~  \phantom\update 
\\ \hline
5b & ~~~~ \begin{minipage}[t]{0.85\textwidth}%
\phantom\moveboundaries~
\end{minipage}
\\ \hline
\rowcolor{yellow!75}
7 & ~~~~ $\phantom\afterupdate $
\\ \hline
\rowcolor{yellow!75}
2 & ~~~~ $ \phantom\invariant  $ 
\\ \hline
 &$\mbox{\color{blue} endwhile} $
\\ \hline \whline
\rowcolor{yellow!75}
2,3 & $ \phantom\invariant \wedge \neg( \phantom\guard )$ 
\\ \hline
\rowcolor{orange!50}
1b & $ \postcondition $ 
\\ \hline
\end{tabular}
}

\newcommand{\FlaWorksheetZero}{
\begin{tabular}{| c | p{0.98\textwidth} |}\hline
Step & $\mbox{\color{blue}Algorithm:~}\routinename$
\\ \hline
\rowcolor{yellow!75}
1a & $ \phantom\precondition $ 
\\ \whline
4 & 
\begin{minipage}[t]{0.9\textwidth}%
$\mbox{\phantom{\partitionings}}$ ~ \\
$\mbox{\color{blue} ~~~where~}$ \phantom{\partitionsizes} 
\end{minipage}
\\ \hline
\rowcolor{yellow!75}
2 & $ \phantom\invariant $ 
\\ \hline
3 &$\mbox{\color{blue}while~} \phantom\guard \mbox{~\color{blue} do}$
\\ \hline 
\rowcolor{yellow!75}
2,3 & ~~~~ $ \phantom\invariant \wedge \phantom \guard$ 
\\ \hline
5a & ~~~~ \begin{minipage}[t]{0.85\textwidth}%
\ifthenelse{\equal{\blocksize}{1}}{}%
{%
\ifthenelse{ \equal{\blocksize}{blank} }{}%
{{\bf Determine block size $ \phantom\blocksize $}\\}%
}
$\mbox{\phantom\repartitionings}$~ \\
$\mbox{\color{blue} ~~~where~}$ \phantom\repartitionsizes
\end{minipage}
\\ \hline
\rowcolor{yellow!75}
6 & ~~~~ $\phantom\beforeupdate $
\\ \hline
8 & ~~~~  \phantom\update 
\\ \hline
5b & ~~~~ \begin{minipage}[t]{0.85\textwidth}%
\phantom\moveboundaries~
\end{minipage}
\\ \hline
\rowcolor{yellow!75}
7 & ~~~~ $\phantom\afterupdate $
\\ \hline
\rowcolor{yellow!75}
2 & ~~~~ $ \phantom\invariant  $ 
\\ \hline
 &$\mbox{\color{blue} endwhile} $
\\ \hline \whline
\rowcolor{yellow!75}
2,3 & $ \phantom\invariant \wedge \neg( \phantom\guard )$ 
\\ \hline
\rowcolor{yellow!75}
1b & $ \postcondition $ 
\\ \hline
\end{tabular}
}

\newcommand{\TBTinitialize}{}
\newcommand{\FlaAlgorithmTBT}{
\begin{tabular}{|l|} \hline
$\mbox{\color{blue}Algorithm:~}\routinename$
\\ \whline
\partitionings \\
$\mbox{\color{blue} ~~~where~}$ \partitionsizes 
\\ 
\TBTinitialize\\
$\mbox{\color{blue}while~} \guard \mbox{~\color{blue} do}$
\\
\ifthenelse{\equal{\blocksize}{1}}{}%
{%
\ifthenelse{ \equal{\blocksize}{blank} }{}%
{~~~~{\bf Determine block size $ \blocksize $}\\}%
}
~~~~ 
\repartitionings \\
~~~$\mbox{\color{blue} ~~~where~}$ \repartitionsizes
\\ \hline
~~~~  \update 
\\ \hline
~~~~ 
\moveboundaries 
\\
$\mbox{\color{blue} endwhile} $
\\ \hline 
\end{tabular}
}

\newcommand{\gemm}{\mbox{\sc gemm}\xspace}

\newcommand{\convn}{\mbox{\sc conv}}
\newcommand{\conv}{\convn\xspace}
\newcommand{\fc}{\mbox{\sc fc}\xspace}

\newcommand{\imcoln}{\mbox{\sc im2col}}
\newcommand{\imcol}{\imcoln\xspace}
\newcommand{\convgemmn}{\mbox{\sc convgemm}}
\newcommand{\convgemm}{\convgemmn\xspace}

\graphicspath{{Figures/}}

\begin{document}

\title{High Performance and Portable Convolution Operators for\\ ARM-based Multicore Processors}

\author{%
Pablo~San~Juan\footnote{Universitat Polit\`ecnica de Val\`encia, Spain. 
          p.sanjuan@upv.es, palonso@upv.es, quintana@disca.upv.es}
\and
Adri\'an Castell\'o\footnote{Universitat Jaume~I, Castell\'on de la Plana, Spain.
                      \{adcastel,dolzm\}@icc.uji.es}
\and
Manuel F. Dolz$^\dag$
\and
Pedro Alonso-Jord\'a$^*$
\and
Enrique S. Quintana-Ort\'{\i}$^*$
}

\date{\today}

\maketitle

\begin{abstract}

The considerable impact of Convolutional Neural Networks on many Artificial Intelligence tasks has led to the development of various high performance algorithms for the convolution operator present in this type of networks. One of these approaches leverages the \imcol transform followed by a general matrix multiplication (\gemm) in order to take advantage of the highly optimized realizations of the \gemm kernel in many linear algebra libraries. The main problems of this approach are 1) the large memory workspace required to  host the intermediate matrices generated by the \imcol transform; and 2) the time to perform the \imcol transform, which is not negligible for complex neural networks. This paper presents a portable  high performance convolution algorithm based on the BLIS realization of the \gemm kernel that avoids the use of the intermediate memory by taking advantage of the BLIS structure. In addition, the proposed algorithm eliminates the cost of the explicit \imcol transform, while maintaining the portability and performance of the underlying realization of \gemm in BLIS.
\end{abstract}

\section{Introduction}
\label{sec:intro}

During the past decade and a half, the use of deep neural networks (DNNs) for machine learning 
(also known as deep learning, or DL), and more specifically convolutional neural networks (CNNs), 
has gained a tremendous momentum, carrying beyond conventional problems in image classification, 
object dectection, speech recognition and neural machine 
translation~\cite{recent-advances-in-deep-learning-for-speech-research-at-microsoft,Krizhevsky:2012:ICD:2999134.2999257,7243232},
to be extended to a myriad of unexplored applications, for example, in
quantum computing, solid state lighting, nanoelectronics and nanomechanics, high throughput screening of new materials,
computer vision in microscopy, radiography and tomography, and 
astrophysics simulation; see~\cite{8114708,Najafabadi2015,DBLP:journals/corr/abs-1802-09941} among many others.

Current CNN models consist of a large number of neuron layers 
that allow to deliver superior accuracy on many artificial intelligence (AI) tasks, at
the cost of a considerable computational cost, both for training and inference~\cite{8114708}.
This cost comes from the CNN being mostly composed of convolutional layers
(\conv), each basically embedding  a high-dimensional convolution operator~\cite{Ma18}.

The high computational cost of the \conv layers can be tackled via 
certain compression techniques (such as 
use of low-rank approximations, quantization/low-precision arithmetic, sparsification, etc.),
which aim to reduce the complexity of the convolution in exchange for a potential degradation in accuracy~\cite{Han15}.
The application of the convolution operator can also be accelerated via optimized implementations of this kernel that
carefully exploit the architecture of modern high performance processors,
such as multicore processors and graphics processing units (GPUs).
On the one hand, when the filters involved in the convolution 
are of size $5 \times 5$ or larger, this kernel is usually realized via the Fast Fourier transform (FFT). 
On the other hand, for smaller (yet more often encountered) filters, 
the operator is cast
in terms of a general matrix multiplication (\gemm)~\cite{chetlur2014cudnn,Geor18,onednn} via the \imcol transform~\cite{Che06}. 
In some cases, the \gemm-based approach can be accelerated employing Winograd's minimal filtering algorithms, 
possibly combined with the Strassen variant of the matrix multiplication~\cite{Lavi16,Zhao18}.
However, this latter strategy can also result in a decay of accuracy of the trained model.

High performance realizations of the convolution operator/\gemm are available in
libraries such as Intel's openDNN/MKL and NVIDIA's cuDNN/cuBLAS, respectively~\cite{onednn,cudnn}.
However, these implementations target Intel/AMD x86 architectures and NVIDIA GPUs, and therefore, they are not 
portable to other architectures. Moreover, except for openDNN, these libraries take a ``black-box'' approach and their contents
cannot be examined nor modified.

The \textit{Basic Linear Algebra Instantiation Software} (BLIS)
is a software framework for rapid development of high-performance dense linear algebra libraries~\cite{BLIS1}. 
BLIS implements the full functionality defined in the 
\textit{Basic Linear Algebra Subprograms} (BLAS) application programming interface (API)~\cite{blas3} featuring several
appealing properties:
\begin{list}{--}{}
\itemsep=2pt\parskip=0pt
\item BLIS is written in Standard C (mainly ISO C90 with a few C99 extensions). 
\item The BLIS code is mostly architecture-independent and, therefore, largely portable. 
      Developing an efficient instance of BLIS for an specific  processor architecture requires an efficient implementation of 
      a small piece of code, known as the micro-kernel, and the selection of a number of cache configuration parameters that
      can be adjusted via an analytical model~\cite{BLIS4}.
\item There exist high performance realizations of the micro-kernel (and tuned selection of the cache configuration parameters) 
      for many different architectures, including
      low-power ARM-based processors~\cite{Catalan2016}. 
\item On a variety of modern multicore processors, 
      BLIS has been shown to deliver sustained high performance~\cite{BLIS2,BLIS3,Catalan2016} that  rivals that of
      commercial libraries, such as Intel's MKL, as well as other open high performance instances of the BLAS,
      such as GotoBLAS~\cite{Goto:2008:AHP,Goto:2008:HPI}, OpenBLAS~\cite{OpenBLAS} and ATLAS~\cite{ATLAS}.
\end{list}

In this paper, \textit{we leverage the open implementation of the \gemm kernel in BLIS to design 
high performance and portable convolution operators for DL inference on  general-purpose multicore processors.}
For this purpose, we modify one of packing routines in the BLIS \gemm kernel to apply the \imcol transform
on-the-fly (that is, during the execution of the matrix multiplication)
on the input tensor for the convolution operator. As a result, our approach features:
\begin{description}
\itemsep=2pt\parskip=0pt
\item[Reduced workspace.] We avoid the explicit assembly of the large-scale matrix that results from applying the \imcol 
      transform to the input tensor, requiring no extra workspace (other than the small buffers that are
      used inside the BLIS \gemm).
\item[High performance.] Our solution
      mimics the performance of the BLIS \gemm, basically eliminating the overhead of the \imcol transform,
      to reduce the execution time of the convolution operator to that of the associated \gemm kernel.
\item[Portability.] The result
      remains as portable as BLIS since our modification of the \gemm kernel does not affect the micro-kernel
      nor the cache configuration parameters.
\end{description}
As an additional contribution of this work, we assess the advantages of our integration of \imcol into the BLIS
\gemm by porting and evaluating the resulting convolution operator on the ARM
quad-core Cortex-A57 processor (ARMv8, 64-bits) that is integrated in
the NVIDIA Jetson TX2 module.

The rest of the paper is organized as follows.
After a survey of related work in the next subsection,
in Section~\ref{sec:blis} we review the 
BLIS approach for the implementation of \gemm, briefly discussing the portability and
multi-threaded parallelization of this kernel. Special attention is paid there to
the packing performed within BLIS, in connection with the layout of the data in memory, as these are two keys to our  approach.
In Section~\ref{sec:conv}
we review the \imcol transform and how to leverage this function to cast a convolution in terms of the matrix multiplication.
We then open Section~\ref{sec:convopt} with a discussion of the problems of such 
straight-forward scheme, proposing an alternative
that embeds the \imcol transform within the BLIS \gemm kernel, yielding a portable, high performance, 
integrated \convgemm operator for multicore processors.
Finally, we evaluate the performance of the new routines on an 
ARM Cortex-A57 processor in Section~\ref{sec:experiments}, and offer some final closing remarks
in Section~\ref{sec:remarks}.

\subsection{Related work}

\paragraph{Direct algorithms.}
Libraries such as NVIDIA's cuDNN, HexagonNN~\cite{Jin19} and Xiaomi's MACE~\cite{mace}
include optimized direct convolution operators for the most frequently encountered filter dimensions and strides,
falling back to default algorithms for other parameter values.
In comparison, Intel's MKL-DNN~\cite{Geor18} employs parameterized architecture-aware just-in-time code generators 
to produce direct optimized convolution routines at runtime. 

NNPACK (Neural Networks PACKage)~\cite{nnpack} also
provides direct implementations of convolutional operators involving large filters (3$\times$3 or 5$\times$5) using 
either Winograd filters or FFT. NNPACK supports many popular deep
learning frameworks (Caffe2, PyTorch, MXNET, etc.) and includes architecture-specific optimizations for
ARMv7, ARMv8, and x86 processors.

\paragraph{Indirect algorithms.}
In contrast with the previous approach, 
\gemm-based algorithms reformulate the convolution in terms of a two-stage (or indirect) \imcoln+\gemm. This allows to
leverage highly optimized realizations of the BLAS, which exists for almost any modern computer platform.
As a result, the \gemm-based approach is now used in all major deep learning frameworks~\cite{Dukh19}.

Facebook's QNNPACK (Quantized NNPACK)~\cite{qnnpack} 
extends NNPACK to perform computations in 8-bit fixed-point precision targeting 
convolution operators which cannot benefit from fast Winograd/FFT-based schemes. 
Similar to our approach, 
QNNPACK follows an indirect approach while 
aiming to eliminate the overhead of the \imcol transform for matrix multiplication libraries.

A few other works have addressed the excessive memory consumption of \gemm-based algorithms
by dividing the matrix multiplication into small kernels~\cite{Cho17,DBLP:journals/corr/abs-1709-03395}. However, the authors of these
works do not consider the combination of their solutions with optimized, architecture-specific realizations of the
\gemm kernel.

In~\cite{Dukh19}, M. Dukhan tackles both the memory and performance issues of the indirect approach.
Concretely, that work proposes to introduce an indirection structure of pointers to the convolution input operand
optimized for the so-called NHWC layout. Unfortunately, the author recognizes that 
1) the algorithm is not expected to be competitive with state-of-the-art patch-building algorithms~\cite{Zhan18} 
due to strided memory access; and
2) his solution has limited applicability for the backward pass of the convolution operator and the Transposed Convolution operator.

\section{Portable and Multi-Threaded \gemm in BLIS}
\label{sec:blis}

\paragraph{General overview.}
Consider the \gemm operation
$C \mathrel{+}= A \cdot B$, where the dimension of the operands are 
$C \rightarrow m \times n$, $A \rightarrow m \times k$, and $B \rightarrow k \times n$.
BLIS adheres to the high-performance taxonomy in GotoBLAS~\cite{Goto:2008:AHP}
to implement this kernel (and any other variant, with transposed $A$ and/or $B$) 
as three nested loops around a \textit{macro-kernel} plus two \textit{packing routines};
see Loops~\textsf{L1}--\textsf{L3} in the \gemm algorithm in \figurename~\ref{fig:gotoblas_gemm}.
In addition, the macro-kernel is implemented in terms of two additional loops around a \textit{%
micro-kernel}; see Loops~\textsf{L4} and~\textsf{L5} in the same figure.
The micro-kernel is encoded as a loop around a 
rank--1 update (that is, and outer product; not explicitly shown in the figure). 
For simplicity, 
we will consider hereafter that $m,n,k$ are integer multiples of $m_c,n_c,k_c$, respectively; and
$m_c,n_c$ are integer multiples of $m_r,n_r$, respectively.

In BLIS, the loop ordering, together with the packing
routines and an appropriate selection of the loop strides $n_c$, $k_c$, $m_c$, $n_r$ and $m_r$ 
(which match the processor cache configuration), orchestrate a regular pattern of data transfers through the 
memory hierarchy~\cite{BLIS1,BLIS4}. In rough detail, given a processor architecture,
the goal is that a $k_c \times n_r$ micro-panel of the buffer $B_c$, say $B_r$, 
and an $m_c \times k_c$ macro-panel of the buffer $A_c$, say $A_r$, 
are streamed into the floating-point units (FPUs) from the L1 and L2 caches, respectively;
while the $k_c \times n_c$ macro-panel $B_c$ resides in the L3 cache (if present).

\paragraph{Portability.}
An appealing property of BLIS is that all routines are encoded in C except, possibly, for the 
rank--1 update inside the micro-kernel, which may be vectorized using either assembly or vector intrinsics~\cite{BLIS1}. 
Furthermore, following the convention for BLAS~\cite{blas3}, the routines for (almost) 
all other Level-3 BLAS are built on top of \gemm.  This enhances portability as, given a ``generic''
(architecture-oblivious) instance of the BLIS \gemm, porting all the BLIS library to a particular processor architecture then
only needs to 
develop an efficient realization of the rank--1 update for the target processor,
and  selecting the proper values for $n_c$, $k_c$, $m_c$, $n_r$ and $m_r$ to the processor cache/memory configuration.

\begin{figure*}[t]
\centering
\begin{minipage}[c]{0.9\textwidth}
\resizebox{\linewidth}{!}{
\begin{tabular}{llll}
\textsf{L1:} &{\bf for} $j_c$ = $0,\ldots,n-1$ {\bf in steps of} $n_c$\\
\textsf{L2:} & \hspace{3ex}  {\bf for} $p_c$ = $0,\ldots,k-1$ {\bf in steps of} $k_c$\\
&\hspace{6ex}           \crb{$B(p_c:p_c+k_c-1,j_c:j_c+n_c-1)$} $\rightarrow \crb{B_c}$ & & // Pack into $B_c$\\
\textsf{L3:} & \hspace{6ex}           {\bf for} $i_c$ = $0,\ldots,m-1$ {\bf in steps of} $m_c$\\
&\hspace{9ex}                     \crr{$A(i_c:i_c+m_c-1,p_c:p_c+k_c-1)$} $\rightarrow \crr{A_c}$ & & // Pack into $A_c$ \\
\cline{2-4} 
\textsf{L4:} &\hspace{9ex} {\bf for} $j_r$ = $0,\ldots,n_c-1$ {\bf in steps of} $n_r$  & & // Macro-kernel\\
\textsf{L5:} &\hspace{12ex}   {\bf for} $i_r$ = $0,\ldots,m_c-1$ {\bf in steps of} $m_r$\\
\cline{2-3}
&\hspace{15ex}             \crg{$C_c(i_r:i_r+m_r-1,j_r:j_r+n_r-1)$} & // Micro-kernel \\
&\hspace{19ex} ~$\mathrel{+}=$     ~\crr{$A_c(i_r:i_r+m_r-1,0:k_c-1)$} \\
&\hspace{19ex} ~~~$\cdot$~~~~\crb{$B_c(0:k_c-1,j_r:j_r+n_r-1)$} \\
\cline{2-3}
\cline{2-4} 
\end{tabular}
}
\end{minipage}
\caption{High performance implementation of \gemm in BLIS. In the code, $C_c \equiv C(i_c:i_c+m_c-1,j_c:j_c+n_c-1)$
is just a notation artifact, introduced to ease the presentation of the algorithm, while $A_c,B_c$ correspond to actual buffers that are involved in data copies.}
\label{fig:gotoblas_gemm}
\end{figure*}

\paragraph{Multi-threaded parallelization.}
BLIS allows to choose, at execution time, which
of the five loops of the \gemm kernel are parallelized.
The multi-threaded parallelization of the BLIS \gemm kernel has been previously analyzed 
for conventional multicore processors~\cite{BLIS2}, modern many-threaded architectures~\cite{BLIS3},
and low-power (asymmetric) ARM-based processors in~\cite{Catalan2016}.
The insights gained from these experimental studies show that
Loop~\textsf{L1} is usually a good candidate 
for multi-socket platforms with on-chip L3 caches; 
Loop~\textsf{L3} should be parallelized when 
each core has its own L2 cache; and
Loops~\textsf{L4} and~\textsf{L5} are convenient choices if the cores share the L2 cache.

\paragraph{Data storage.}
Hereafter, unless otherwise explicitly stated,
we adhere to the Fortran convention that dictates the column-major order storage for matrices. 
This implies that, for example, the entries of the 2D array (i.e., matrix) 
$C \rightarrow \crb{m} \times \crr{n}$, are arranged in
consecutive positions in memory as
\[
\begin{array}{l}
    \underbrace{C[\crb{0}][\crr{0}],~C[\crb{1}][\crr{0}],\ldots, C[\crb{m-1}][\crr{0}]}_{\textrm{1st column of }C},
\quad
    \underbrace{C[\crb{0}][\crr{1}],~C[\crb{1}][\crr{1}],\ldots, C[\crb{m-1}][\crr{1}]}_{\textrm{2nd column of }C},\ldots,\\[0.3in]
    ~~~~\underbrace{C[\crb{0}][\crr{n-1}],~C[\crb{1}][\crr{n-1}],\ldots,C[\crb{m-1}][\crr{n-1}]}_{\textrm{Last column of }C}.
\end{array}
\]
Note that BLAS follows the Fortran matrix storage convention and, therefore, this is necessary to 
be able to invoke the \gemm kernel.

\paragraph{The packing routines.}
The purpose of these routines is 
to arrange the elements of
$A$ and $B$ into $A_c$ and $B_c$, respectively,
so that the elements of the $A_c$ and $B_c$ buffers
will be accessed with unit stride when executing the micro-kernel~\cite{henryPacking}.
(An additional benefit of packing is that $A_c$ and
$B_c$ are preloaded into certain cache levels of the memory hierarchy,
reducing the time to access the elements of these buffers when using them to update a micro-tile of $C$.)

\begin{figure}[tbp!]
\centering
\begin{tabular}{c}
\includegraphics[width=0.5\textwidth]{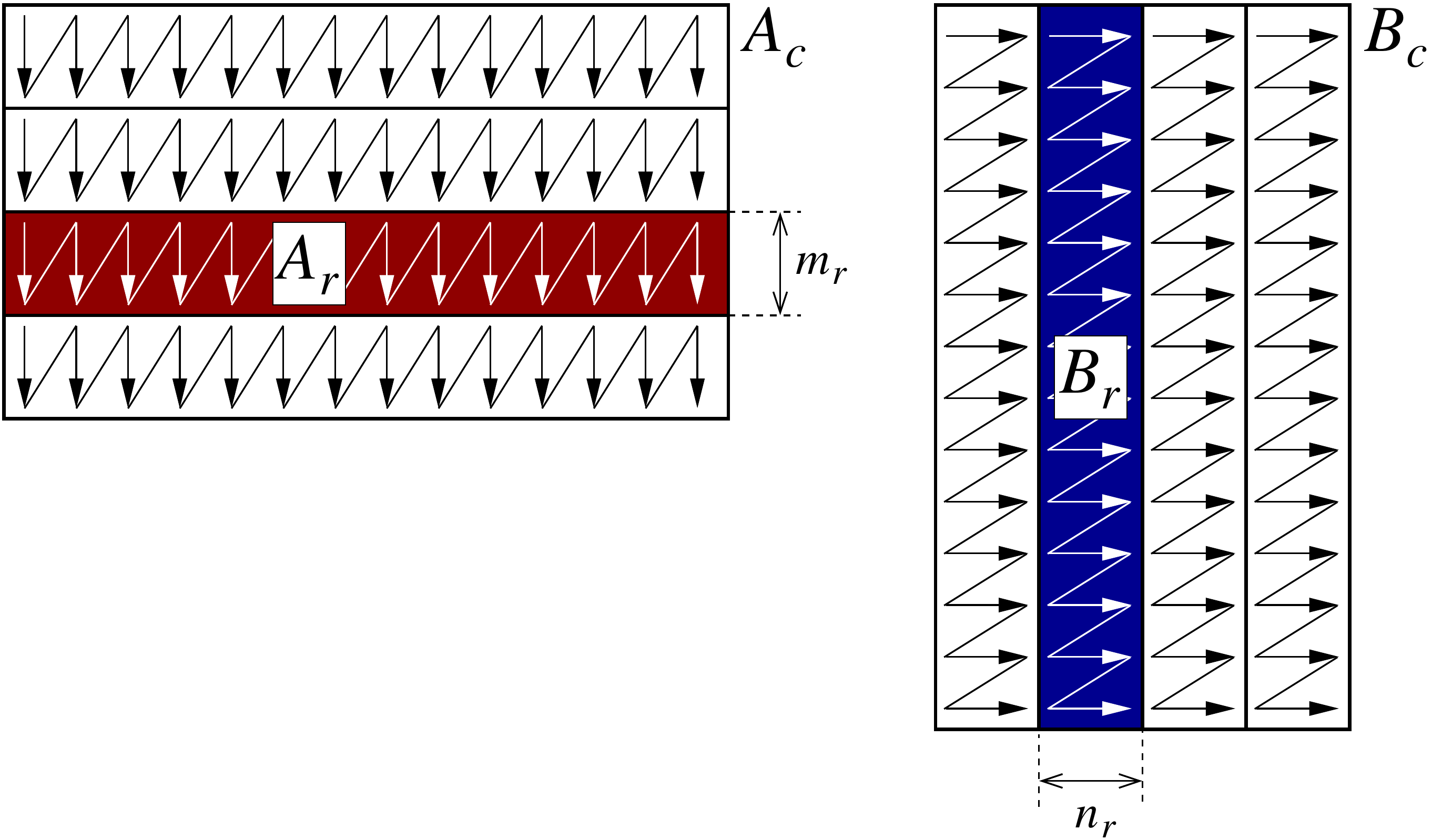}
\end{tabular}
\caption{Packing in the BLIS and GotoBLAS implementations of \gemm. The arrows indicate the linear layout of the 
         elements in the memory: column-major for $A_r$ and row-major for $B_r$.}
\label{fig:blis_packing}
\end{figure}
 
The packing routines proceed to copy and compact the data of the input operands as follows.
In the packing routine for $A_c$,
each $m_c \times k_c $ block of $A$ is packed into the $A_c$ buffer with its elements organized as micro-panels of 
size $m_r \times k_c$; furthermore, within each micro-panel of $A_c$, the elements are stored in
column-major order. Also, each $k_c \times n_c$ block of $B$ is packed into
$B_c$, with its the elements arranged into micro-panels of
size $k_c \times n_r$; and each micro-panel stored into
row-major order; see Figure~\ref{fig:blis_packing} and the algorithm in Figure~\ref{fig:packB}.

\begin{figure}[tb!]
\centering
\footnotesize
\begin{tabular}{l@{\hspace{6pt}}l@{}}
\\ \textsf{L1:} &\hspace{0ex}  {\bf for} $j_r=0,\ldots,n_c-1$ {\bf in steps of} $n_r$ 
\\              &\hspace{4ex} $i=0$
\\ \textsf{L2:} &\hspace{4ex}  {\bf for} $p_s=0,\ldots,k_c-1$ 
\\ \textsf{L3:} &\hspace{8ex}  {\bf for} $j_s=0,\ldots,n_r-1$
\\              &\hspace{12ex} $B_c[i][j_r]  = B[p_c+p_s][j_c+j_r+j_s]$
\\              &\hspace{12ex} $i = i +1$
\end{tabular}
\caption{Algorithm for packing $B$ into $B_c$. 
         The indices $p_c$ and $j_c$ correspond to the
         coordinates of the top-left entry for the block of matrix $B$ that is packed; see Figure~\ref{fig:gotoblas_gemm}.
         Matrix $B$ is maintained in column-major order. Each micro-panel $B_r$ within the buffer $B_c$ is arranged
         in row-major order, as expected by the BLIS micro-kernel; see Figure~\ref{fig:blis_packing}. This
         is attained by viewing $B_c$ as an $(k_c \cdot n_r) \times (n_c/n_r)$ buffer, where each column contains an
         entire micro-panel in row-major order.}
\label{fig:packB}
\end{figure}

Let us consider the overhead introduced by the data copies necessary to perform the packing. Consider, for example,
the packing for $B_c$. 
In principle,  packing this buffer requires $k_c \cdot n_c$ memory accesses, to read the elements of matrix
$B$ (from the memory) 
and write them into the appropriate positions of the buffer (in principle, in the L3 cache, if there is one).
Each buffer is then re-utilized for the (floating-point) operations embraced by Loop~\textsf{L3} of the \gemm
kernel (see Figure~\ref{fig:gotoblas_gemm}),
which amount to 
$\frac{m}{m_c} \cdot
 \frac{n_c}{n_r} \cdot
 \frac{m_c}{m_r} \cdot
 2(m_r n_r k_c) = 2 m n_c k_c$ flops.
Thus, provided $m$ is large enough, the cost of the packing for $B_c$ is negligible compared with the amount of flops
performed inside Loop~\textsf{L3}. A similar reasoning applies to the overhead due to the packing for $A_c$.

As we will expose in the next section, the packing
routines are particularly important for our implementation of the convolution operator.

\section{Indirect Convolution via Explicit \imcoln+\gemm}
\label{sec:conv}

\paragraph{Convolution operator.}
Consider a \conv layer, 
appearing during inference with a DNN model, that comprises a convolution operator
consisting of $k_{n}$ filters (or kernels) of dimension $ k_h \times k_w \times c_{i}$ each.
Assume the layer receives $b$ tensor 
inputs of dimension  $h_{i} \times w_{i} \times c_{i}$ each; and produces
$b$ tensor outputs of size  $h_{o} \times w_{o} \times k_{n}$ each. 
(The parameter $b$ is also often referred to as the batch size.)
Then, each of the $k_n$ individual filters
in this layer combines a (sub)tensor of the inputs, with the same dimension as the filter,
to produce a single scalar value (entry) in one of the $k_n$ outputs.
By repeatedly applying the filter
to the whole input,  in a sliding window manner (with a certain stride $s$),
the convolution operator produces the complete entries of this single output; see~\cite{8114708}. Assuming 
a padding $p$ along dimensions $h_i$ and $w_i$, the output dimensions become 
$h_o=\lfloor (h_{i}-k_h + 2 p)/s +1\rfloor$ and
$w_o=\lfloor (w_{i}-k_w + 2 p)/s +1\rfloor$. 

The algorithm in Figure~\ref{fig:conv} provides a \textit{direct realization} of a 
convolution operator
         $O = \textrm{\conv}(F,\,I)$,
where
$I \rightarrow h_i \times w_i \times c_{i} \times b$ corresponds to the input tensor,
$F \rightarrow k_{n} \times k_{h} \times k_w \times c_i$ denotes the filters, and
$O \rightarrow k_n \times h_o \times w_o \times b$ is the output tensor.

\begin{figure}[tb!]
\centering
\footnotesize
\begin{tabular}{l@{\hspace{6pt}}l@{}}
\\ \textsf{L1:} &\hspace{0ex}  {\bf for} $i_b=0,\ldots,b-1$
\\ \textsf{L2:} &\hspace{4ex}  {\bf for} $i_c=0,\ldots,c_i-1$
\\ \textsf{L3:} &\hspace{8ex}  {\bf for} $i_w=0,\ldots,w_{o}-1$
\\ \textsf{L4:} &\hspace{12ex} {\bf for} $i_h=0,\ldots,h_{o}-1$
\\ \textsf{L5:} &\hspace{16ex}  {\bf for} $i_{kw}=0,\ldots,k_w-1$
\\ \textsf{L6:} &\hspace{20ex}  {\bf for} $i_{kh}=0,\ldots,k_h-1$
\\ \textsf{L7:} &\hspace{24ex} {\bf for} $i_k=0,\ldots,k_n-1$
\\              &\hspace{28ex} $O[i_k][i_h][i_w][i_b]$
$+=$ \hspace{0ex} $F[i_k][i_{kh}][i_{kw}][i_{c}]$
~$\cdot$~ 
$I[s \cdot i_h+i_{kh}][i_w \cdot s +i_{kw}][i_c][i_b]$
\end{tabular}
\caption{Direct algorithm for the application of the convolution operator $O = \textrm{\conv}(F,\,I)$.}
\label{fig:conv}
\end{figure}

\paragraph{Tensor data storage.} 
A tensor generalizes the concept of a matrix to that of a multidimensional array. Note though 
that, from the physical point of view, the tensor entries are still arranged as a linear array in memory. 
Here, we generalize the Fortran convention of 
column-major order to consider that, unless explicitly stated otherwise, 
the entries of the tensors are stored in consecutive positions
in memory starting from the leftmost indices.
This implies that, for example, if the tensor
$O \rightarrow k_n \times h_o \times w_o \times b$
is stored into an 4D array $O[\crb{k_n}][\crr{h_o}][\crg{w_o}][\crc{b}]$, then its entries are consecutively 
arranged in memory as
\[
\begin{array}{l}
    O[\crb{0}][\crr{0}][\crg{0}][\crc{0}],~O[\crb{1}][\crr{0}][\crg{0}][\crc{0}],\ldots, O[\crb{k_n-1]}[\crr{0}][\crg{0}][\crc{0}],~\\[0.1in]
    ~~~~O[\crb{0}][\crr{1}][\crg{0}][\crc{0}],~O[\crb{1}][\crr{1}][\crg{0}][\crc{0}],\ldots, O[\crb{k_n-1}][\crr{1}][\crg{0}][\crc{0}],\\[0.1in]
    ~~~~~~~~~~\ddots\\[0.1in]
    ~~~~~~~~O[\crb{0}][\crr{h_o-1}][\crg{0}][\crc{0}],~O[\crb{1}][\crr{h_o-1}][\crg{0}][\crc{0}],\ldots, O[\crb{k_n-1}][\crr{h_o-1}][\crg{0}][\crc{0}],\\[0.1in]
    ~~~~~~~~~~~~O[\crb{0}][\crr{0}][\crg{1}][\crc{0}],~O[\crb{1}][\crr{0}][\crg{1}][\crc{0}],\ldots, O[\crb{k_n-1}][\crr{0}][\crg{1}][\crc{0}],\ldots,\\[0.1in]
    ~~~~~~~~~~~~~~~~O[\crb{k_n-1}][\crr{h_o-1}][\crg{w_o-1}][\crc{b-1}].
\end{array}
\]

\paragraph{Indirect convolution and the \imcol transform.}
On modern computer architectures, the performance of
the direct realization of the convolution operator given in Figure~\ref{fig:conv} is limited
by the memory bandwidth and, therefore, delivers only a fraction of the processor peak floating-point throughput.
In practice, higher performance can be attained via an 
\textit{indirect (or \gemm-based) approach} that casts this operator in terms of a matrix multiplication via the
\imcol transform~\cite{Che06}.
Concretely, the algorithm in Figure~\ref{fig:im2col}
shows how to transform the input tensor $I$ into an 
augmented matrix $\hat{B}$. With this transform, 
the output of the application of the convolution can be simply obtained from the \gemm $\hat{C} = \hat{A}\cdot \hat{B}$,
where
$\hat{C} \equiv O \rightarrow k_n \times (h_o \cdot w_o \cdot b)$ is the output tensor (viewed as an $m \times n$ matrix,
with $m=k_n$ and $n=(h_o \cdot w_o \cdot b)$);
$\hat{A} \equiv F \rightarrow k_n \times (k_h \cdot k_w \cdot c_i)$ contains the kernels; and
$\hat{B} \rightarrow (k_h \cdot k_w \cdot c_i) \times (h_o \cdot w_o \cdot b)$
is the result from applying the \imcol transform to the input tensor~$I$.

\begin{figure}[tb!]
\centering
\footnotesize
\begin{tabular}{l@{\hspace{6pt}}l@{}}
   \textsf{L1:} &\hspace{0ex}  {\bf for} $i_b=0,\ldots,b-1$
\\ \textsf{L2:} &\hspace{4ex}  {\bf for} $i_c=0,\ldots,c_i-1$
\\ \textsf{L3:} &\hspace{8ex}  {\bf for} $i_w=0,\ldots,w_{o}-1$
\\ \textsf{L4:} &\hspace{12ex}  {\bf for} $i_h=0,\ldots,h_{o}-1$
\\              &\hspace{16ex} $c =   i_{h}+ i_w \cdot h_i + i_b \cdot w_i \cdot h_i$
\\ \textsf{L5:} &\hspace{16ex}  {\bf for} $i_{kw}=0,\ldots,k_w-1$ 
\\ \textsf{L6:} &\hspace{20ex}  {\bf for} $i_{kh}=0,\ldots,w_h-1$ 
\\              &\hspace{24ex} $r ~~~~~~~=   i_{kh} + i_{kw} \cdot k_h + i_c \cdot k_w \cdot k_h$
\\              &\hspace{24ex} $\hat{B}[r][c] = I[i_h \cdot s + i_k{h}][i_w \cdot s + i_{kw}][i_c][i_b]$
\end{tabular}
\caption{Algorithm for the \imcol transformation. The actual implementation moves some of the loop invariants
         inside Loops~\textsf{L4} and~\textsf{L6} to reduce the indexing arithmetic overhead.
         For simplicity, this is not shown in the algorithm.}
\label{fig:im2col}
\end{figure}

\section{Optimized Indirect Convolutions via Integration of \imcol into \gemm}
\label{sec:convopt}

There are two problems with the indirect (two-stage) procedure described in Section~\ref{sec:conv} that performs the convolution
as a sequence of an explicit \imcol transform followed by a call to the \gemm kernel:
\begin{description}
\itemsep=2pt\parskip=0pt
\item[P1.] Starting from an input tensor $I$ of dimension
$ h_i \times w_i \times c_{i} \times b $,
the \imcol transforms creates an augmented matrix $\hat{B}$ of size
$(k_h \cdot k_w \cdot c_i) \times (h_o \cdot w_o \cdot b)$. Assuming
$h_i,w_i \approx h_o,w_o$, this requires a workspace that is $k_h \cdot k_w$ times larger than the original input tensor.
For current CNNs, with many layers, even when using small $3\times 3$ filters, this can easily exceed the memory capacity of
the system.
\item[P2.] 
      On modern high performance processors, when using a realization of the \gemm kernel that
      is highly optimized, the overhead due to the copy and replication
      required by the \imcol transform in general becomes
      ``visible'' and reduces the performance of the global (explicit) \imcol+BLIS \gemm process.
\end{description}

To tackle both problems, we propose a solution that integrates the 
\imcol transform into the packing of $\hat{B}$ onto the buffer $B_c$.
In other words, during the execution of the \gemm kernel, the buffer $B_c$ is directly assembled
from the contents of the input tensor $I$ (instead of using the augmented matrix $\hat{B}$, which is never created).
\textit{In the following, we will refer to our solution as an indirect convolution via a \convgemm operator.}
We can now justify the contributions listed in the introduction of this work (see Section~\ref{sec:intro}):
\begin{description}
\itemsep=2pt\parskip=0pt
\item[Reduced workspace.] We avoid the use of the large workspace present in the two-step procedure (problem \textsf{\bf P1}), 
      as the only 
     ``additional'' storage that is needed is the buffer for $B_c$, which
     is already necessary in the BLIS \gemm kernel. 
\item[High performance.] Furthermore, as argued during the discussion of the packing in Section~\ref{sec:blis},
       the memory access costs introduced by the packing of $B_c$ is well amortized with the flops that are performed in the 
       innermost loops and, therefore, the overhead can be considered negligible (problem \textsf{\bf P2}).
\item[Portability.] The approach has the additional advantage that the only change that is needed to the BLIS
\gemm is to replace the original packing routine with a procedure that reads (and packs) the second input operand to the matrix
multiplication directly from the input tensor. There is no need to modify the routine that performs the packing 
with $\hat{A}$. More importantly, there is no need to change the micro-kernel, which enhances the portability of our solution:
the only part that is different is written in C and depends on a small number of architecture-dependent parameters
that are adjusted during the process of porting BLIS.
The parameters that define the filter dimensions are ``embedded'' within the dimensions of the resulting matrix and,
therefore, require no specific optimization.
\end{description}

The algorithm in Figure~\ref{fig:packI} 
illustrates how to pack the corresponding entries of the input tensor $I$ into
the buffer $B_c$ during the execution of the BLIS \gemm kernel 
in Figure~\ref{fig:gotoblas_gemm} while, simultaneously, performing the implicit \imcol transform. 
The algorithm packs the $k_c \times n_c $ block of  matrix $\hat{B}$ starting at row $p_c$ and column $k_c$ into the buffer $B_c$, reading the corresponding entries directly from the input tensor $I$. 
As a result, the output matrix comprises the sought-after convolution:
\[
O = \textrm{\conv}(F,\,I)  \quad \equiv \quad
\hat{C} = \hat{A} \cdot \hat{B} \quad \equiv  \quad
\hat{C} = \hat{A} \cdot \imcol(I),
\]
where
$\hat{C} \equiv O$ and
$\hat{A} \equiv F$.
The actual implementation of this algorithm eliminates some of the loop invariants and integer arithmetic to reduce the overhead. 
Concretely, the computation of the indices $i_c,i_{kw},i_{kh},i_b,i_w,i_h$ is performed outside 
the loops and then properly updated during the iterations to avoid the high cost of the integer divisions and modulo
operations (that is, the remainder of the integer division, abbreviated in the presentation as mod). 
The algorithm is shown in this basic form to improve readability.

\begin{figure}[tb!]
\centering
\footnotesize
\begin{tabular}{l@{\hspace{6pt}}l@{}}
\\ \textsf{L1:} &\hspace{0ex}  {\bf for} $j_r=0,\ldots,n_c-1$ {\bf in steps of} $n_r$
\\              &\hspace{4ex} $i=0$
\\ \textsf{L2:} &\hspace{4ex}  {\bf for} $p_s=0,\ldots,k_c-1$
\\              &\hspace{8ex} $i_c = (p_c + p_s) / (k_h \cdot k_w) $
\\              &\hspace{8ex} $i_{kw} = ((p_c + p_s) \bmod{(k_h \cdot k_w)})/ k_h $
\\              &\hspace{8ex} $i_{kh} = ((p_c + p_s) \bmod{(k_h \cdot k_w)})\bmod{k_h} $
\\ \textsf{L3:} &\hspace{8ex}  {\bf for} $j_s=0,\ldots,n_r-1$
\\              &\hspace{12ex} $i_b = (j_c + j_r + j_s) / (h_o \cdot w_o) $
\\              &\hspace{12ex} $i_w = ((j_c + j_r + j_s) \bmod{(h_o \cdot w_o)})/ h_o $
\\              &\hspace{12ex} $i_h = ((j_c + j_r +j_s) \bmod{(h_o \cdot w_o)})\bmod{h_o} $
\\              &\hspace{12ex} $B_c[i][j_r]  = I[i_{kh} + i_h \cdot s][i_{kw} + i_w \cdot s][i_c][i_b]$
\\              &\hspace{12ex} $i = i +1$
\end{tabular}
\caption{Algorithm for packing $I$ into $B_c$. The indices $p_c$ and $j_c$ correspond to the
         coordinates of the top-left entry for the block of matrix $\hat{B}$ that is packed; see Figure~\ref{fig:gotoblas_gemm}.%
}

\label{fig:packI}
\end{figure}

\section{Performance Evaluation}
\label{sec:experiments}

In this section, we assess the performance of our \convgemm approach (that is, \imcol integrated into the BLIS \gemm) 
against the baseline counterpart that
explicitly assembles the extended input activation matrix and then performs the augmented \gemm. 
As described next, 
for this evaluation
we target a high performance
ARM processor present in a low-power embedded system, and perform the analysis by simulating the inference stage of three
representative state-of-the-art CNNs.
%
The source of all codes employed for the evaluation, including the \convgemm implementation, is publicly available in a git repository~\cite{gitRepo}. 

\subsection{Configuration}
 The evaluation presented in this paper was executed on an NVIDIA Jetson TX2~\cite{jetson} platform, 
 which integrates 
     an ARM quad-core Cortex-A57, 
     an NVIDIA dual-core Denver, 
     an NVIDIA 256-CUDA core Pascal GPU, and
     8 GiB of main memory. 
 The results reported next were obtained in the ARM Cortex-A57 only, due to the wide spread of this architecture and 
 the availability of optimized high performance linear algebra libraries for this processor.
 On the software side, the experiments were conducted using the Linux distribution
     Ubuntu 18.04.4,
     the GNU compiler gcc 7.5.0, and
     BLIS 0.6.0.
  
 As the evaluation targets inference with CNNs, all the experiments employed (IEEE) simple precision arithmetic. 
In general, the inference process does not benefit from the use of double precision arithmetic, 
 and a reduced precision format (floating point single or half, or even fixed point) 
 is often preferred in order to improve performance and/or reduce energy consumption.
 BLIS provides a single-precision instance of the BLAS optimized for the ARM Cortex-A57 which features an optimized 
 micro-kernel with $m_r \times n_r = 8 \times 12$, and sets the following cache configuration values: 
 $n_c=3072$, $k_c=640$, and $m_c=120$. 
  The algorithm paralellizes loop \textsf{L4} of Figure~\ref{fig:gotoblas_gemm}
  and the outermost loop of the packing of $A$ using OpenMP~\cite{openmp08}. 
  For the \convgemm, we also parallelize
  loop \textsf{L1} of Figure \ref{fig:packI}.
  The counterpart with an explicit \imcol parallelizes loop \textsf{L2}  of   Figure~\ref{fig:im2col}.

  \subsection{Inference simulator}\label{sec:implDetails}

  In order to tackle the complex software stack required for executing CNNs, we have employed an inference simulator that performs the major computational stages of the convolutional layers encountered during the inference of CNN models. 
  For the baseline case,  we emulate this behavior by executing a sequence of explicit \imcoln+\gemm pairs, of the dimensions appearing in consecutive layers of the neural network. 
  Our optimized alternative instead executes the specialized \convgemm kernel (of
  the dimensions dictated by the CNN model). 
  In both cases, the simulator reads the CNN configuration parameters for a certain model from an input file, 
  accepting the batch size (number of input samples simultaneously processed per inference process) as an additional parameter. 
  The simulator then allocates memory buffers for all required matrices using the maximum size of each matrix from among the matrix sizes required by each layer in the model, and performs a full model evaluation for each batch size in the specified range.
During inference, the output of a certain layer is basically the input data of the next layer. Our code mimics this
behaviour by using buffer swapping. In this way, we simulate more accurately the real data movements that take place across the 
cache hierarchy during the inference stage. 

  The simulator repeatedly executes the computational operations till a certain time threshold is attained, 
  and then divides the total wall-time by the number of repetitions to avoid system load variability in the measurements. 
\subsection{DNN Models}\label{sec:models}

 We have applied the simulator to study the benefits of the optimized indirect \convgemm algorithm using three 
 representative CNN models: AlexNet~\cite{Krizhevsky:2012:ICD:2999134.2999257}, VGG16~\cite{simonyan2014very}, and ResNet50~\cite{he2016deep}.\footnote{The models adhere to the specifications defined in Google's TensorFlow benchmarks suite.} The former model was selected because of its simplicity, which facilitates an easier interpretation of the results. 
 The remaining two models were chosen because of their more complex structures and notable computational requirements. Table~\ref{tab:models} summarizes the number and type of layers for each model as well as 
 the extra memory consumption required by the explicit \imcol transform. This later parameter represents the maximum memory needed to hold the largest intermediate matrix assembled by the explicit \imcol transform when executing each model. This is a 
 key parameter because it may constrain the use of the explicit \imcoln+\gemm approach for many CNN model+platform pairs due
 to insufficient memory capacity. 
 \textit{Remember that our optimized algorithm with \convgemm saves this extra space by avoiding the explicit creation of the intermediate matrices.}

\begin{table}
\centering
\begin{tabular}{|l||c|c|c|c|c|}
\hline
Model & \fc & \conv & \textsc{Pool} & Total & Memory consumption for \imcol (MiB)\\ \hline\hline
AlexNet & 3 & ~5 & 3 & 11 & $~15.87 \, b$ \\ 
ResNet50 & 1 & 53 & 1 & 55 & $~13.05 \, b$ \\ 
VGG16 & 3 & 13 & 5 & 21 & $110.25 \, b$\\ \hline
\end{tabular}
\caption{Number and type of layers in the target CNN models 
         and memory required by the explicit \imcol transform as a function of the  batch size $b$.}
\label{tab:models}
\end{table}

Table~\ref{tab:alexnet} details the configuration of the \conv layers for the AlexNet model. Concretely, the table 
displays the number of neurons (represented by the dimensions of the input data); the kernel specifications
(number of kernels, their height and width, and their number input channels); 
and the dimensions of the \gemm product, when applying the indirect convolution, for each layer of that type. 

\begin{table}
\centering
\begin{tabular}{|c||c|c|c|c|c|}
\hline
Layer & Type  & Neurons & Kernels &  \gemm dimensions               \\
 & & $(h_i \times w_i \times c_i \times b)$ & $(k_n \times k_h \times k_w \times c_i)$  & $(m\times n \times k)$ \\
\hline \hline
2  & \conv           & $224 \times 224 \times ~~3 \times b$ & $~64 \times 11 \times 11 \times ~~3$  & $~64\times 2916\,b \times ~363$\\ 
4  & \conv           & $~55 \times ~55 \times ~64 \times b$ & $192 \times ~5 \times ~5 \times ~64$  & $192\times 2601\,b \times 1600$\\ 
6  & \conv           & $~27 \times ~27 \times 192 \times b$ & $384 \times ~3 \times ~3 \times 192$  & $384\times ~625\,b \times 1728$ \\ 

7  & \conv           & $~13 \times ~13 \times 384 \times b$ & $384 \times ~3 \times ~3 \times 384$  & $384\times ~121\,b \times 3456$\\ 
8  & \conv           & $~13 \times ~13 \times 384 \times b$ & $384 \times ~3 \times ~3 \times 384$  & $256\times ~121\,b \times 3456$\\ 
\hline 
\end{tabular}
\caption{Specification for the \conv layers appearing in the AlexNet CNN model as a function of the batch size $b$.}
\label{tab:alexnet}
\end{table}

\subsection{Experimental results}

In this subsection we report the results obtained with the simulator applied to simulate the inference process for
the three selected CNN models. In these experiments, we compare the execution time of the models 
with either 1) an \imcol operation followed by the \gemm on the augmented matrix (explicit \imcoln+\gemm); or 
2) an \imcol performed
on-the-fly with the \gemm (referred to as \convgemmn). To better understand the source of the 
observed differences, in the comparison we also include 3) the cost of the \gemm operations
without (the overhead caused by) the \imcol transforms; and 4) the separate cost of the latter.
Note that, as our ultimate goal is to hide completely the cost of the \imcol transform inside the \gemm operation, 
the performance reference for our \convgemm routine is to match the execution time/performance rate of the
standalone \gemm kernel. 

\begin{figure}[p]
  \begin{subfigure}{0.5\textwidth}
  \centering
  \includegraphics[width=1.0\textwidth]{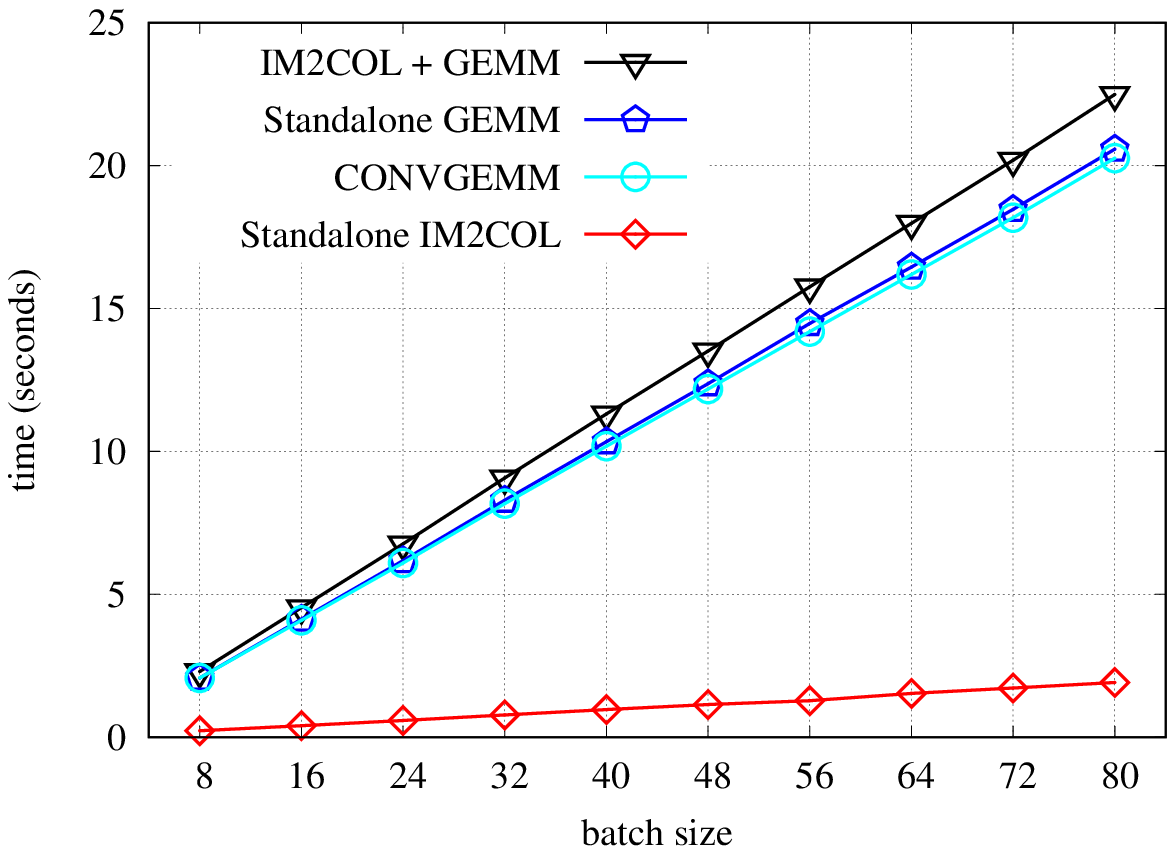}
  \caption{Execution time, AlexNet, 1 core}
  \label{fig:results1ta}
  \end{subfigure}
  \begin{subfigure}{0.5\textwidth}
  \centering
  \includegraphics[width=1.0\textwidth]{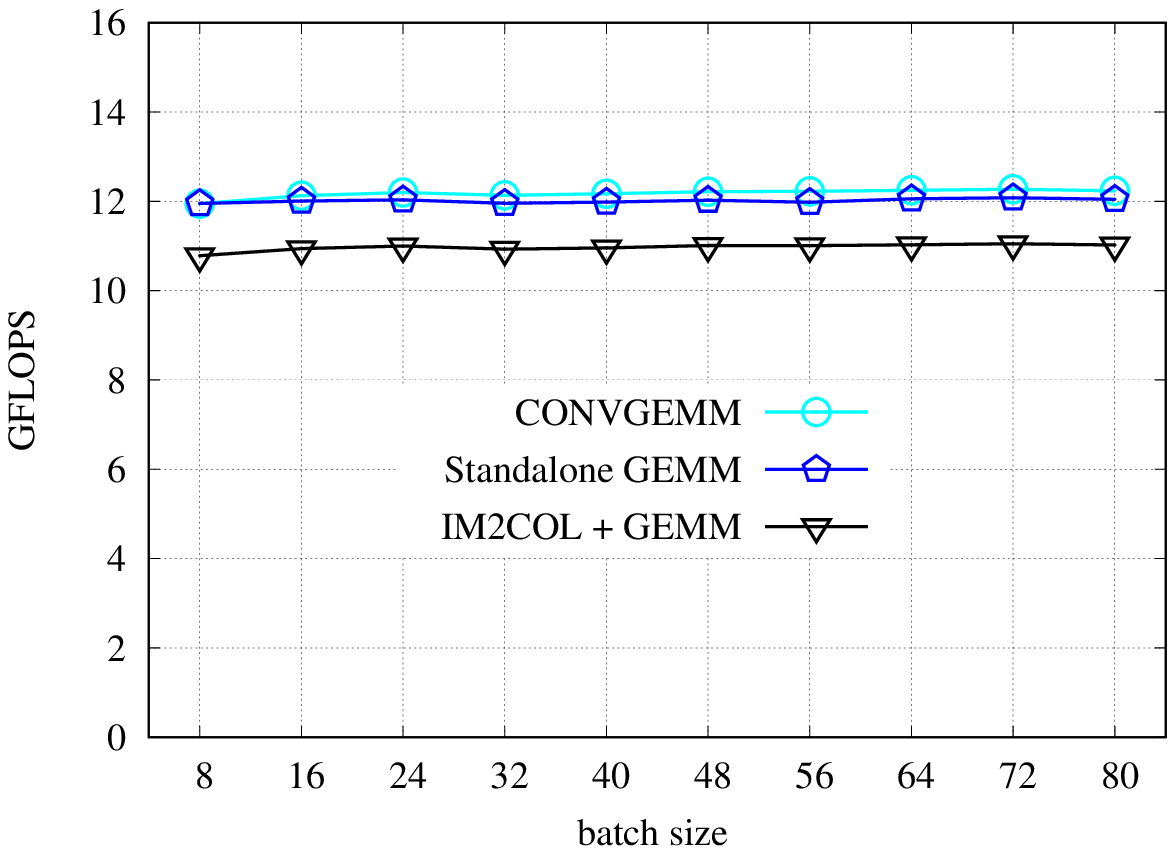}
  \caption{GFLOPS, AlexNet, 1 core}
  \label{fig:results1tb}
  \end{subfigure} \\

  \begin{subfigure}{0.5\textwidth}
  \centering
  \includegraphics[width=1.0\textwidth]{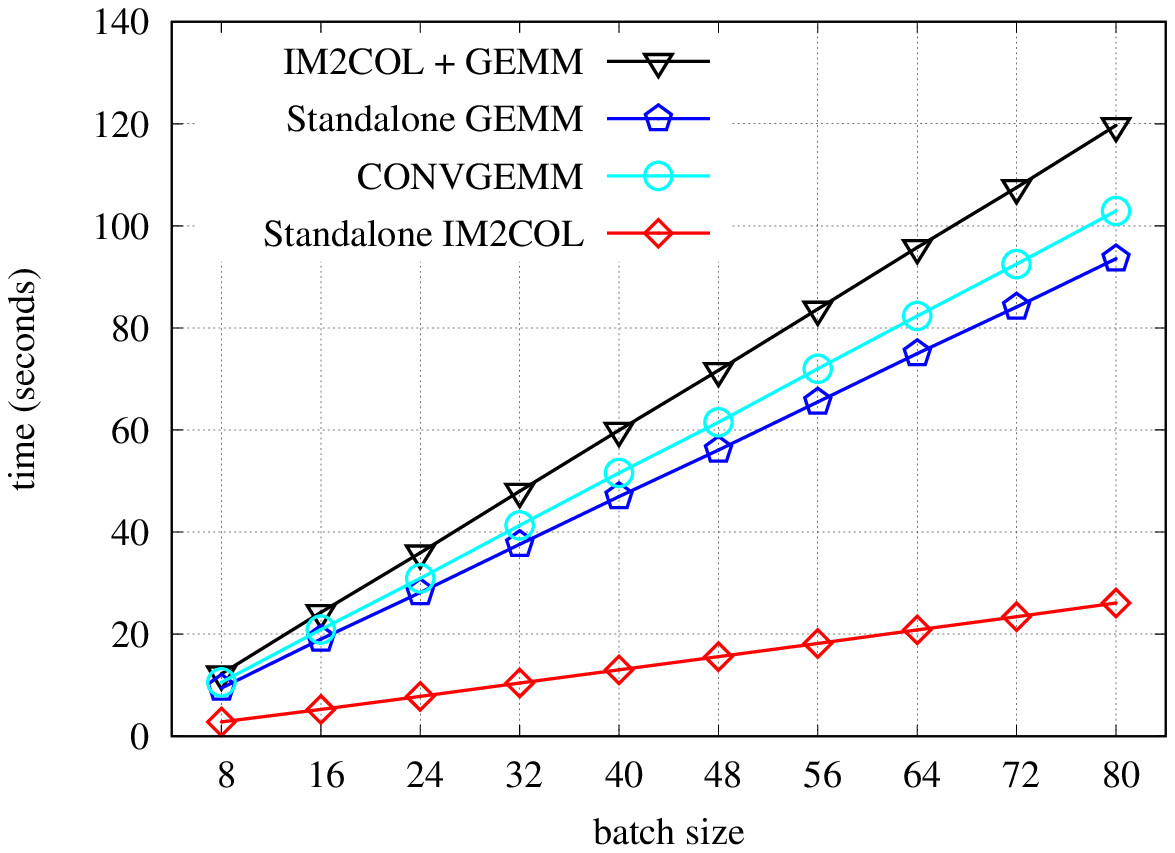}
  \caption{Execution time, ResNet50, 1 core}
  \label{fig:results1tc}
  \end{subfigure}
  \begin{subfigure}{0.5\textwidth}
  \centering
  \includegraphics[width=1.0\textwidth]{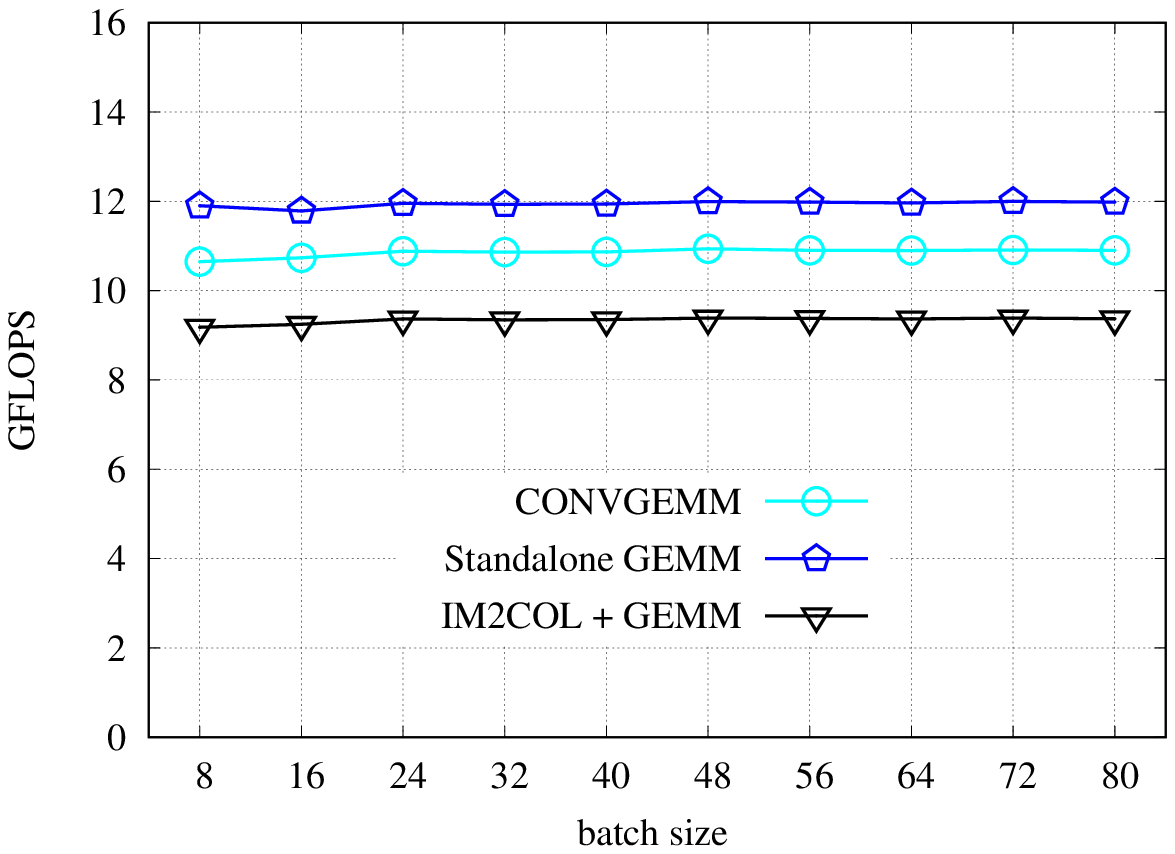}
  \caption{GFLOPS, ResNet50, 1 core}
  \label{fig:results1td}
  \end{subfigure} \\

  \begin{subfigure}{0.5\textwidth}
  \centering
  \includegraphics[width=1.0\textwidth]{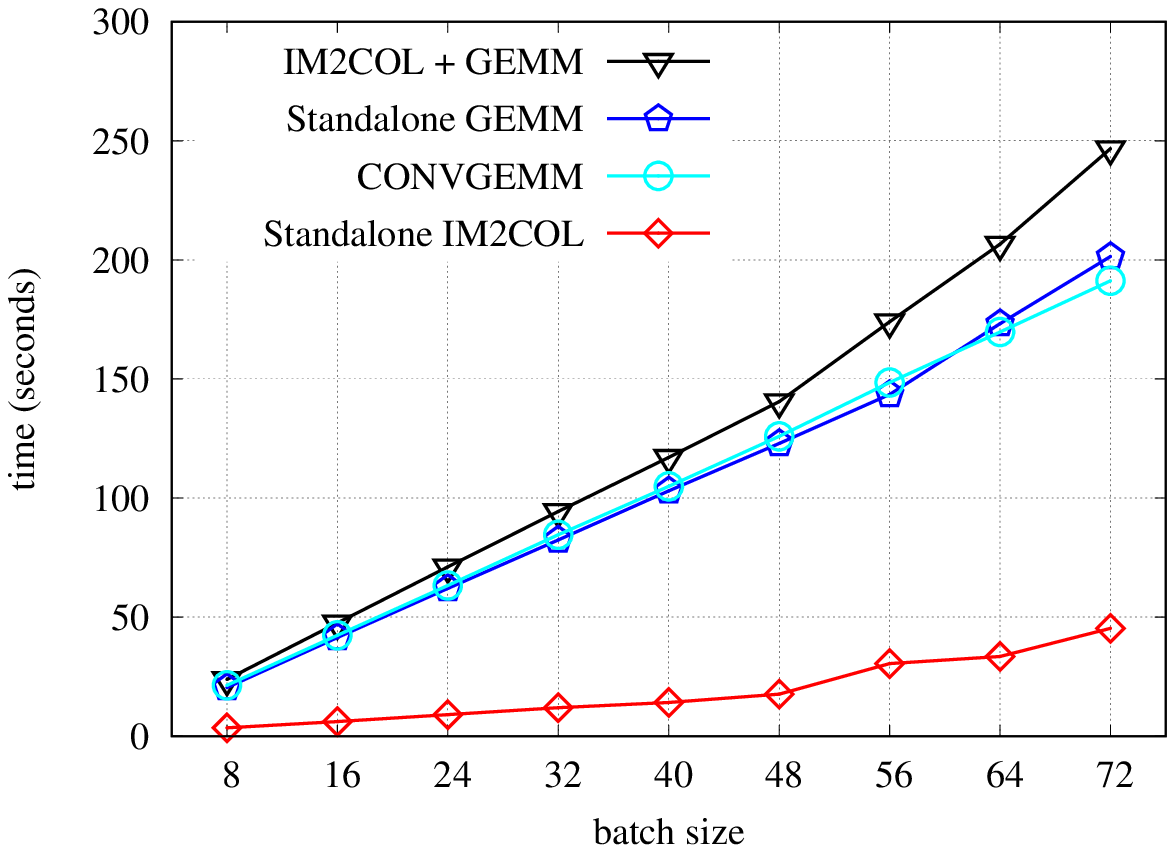}
  \caption{Execution time, VGG16, 1 core}
  \label{fig:results1te}
  \end{subfigure}
  \begin{subfigure}{0.5\textwidth}
  \centering
  \includegraphics[width=1.0\textwidth]{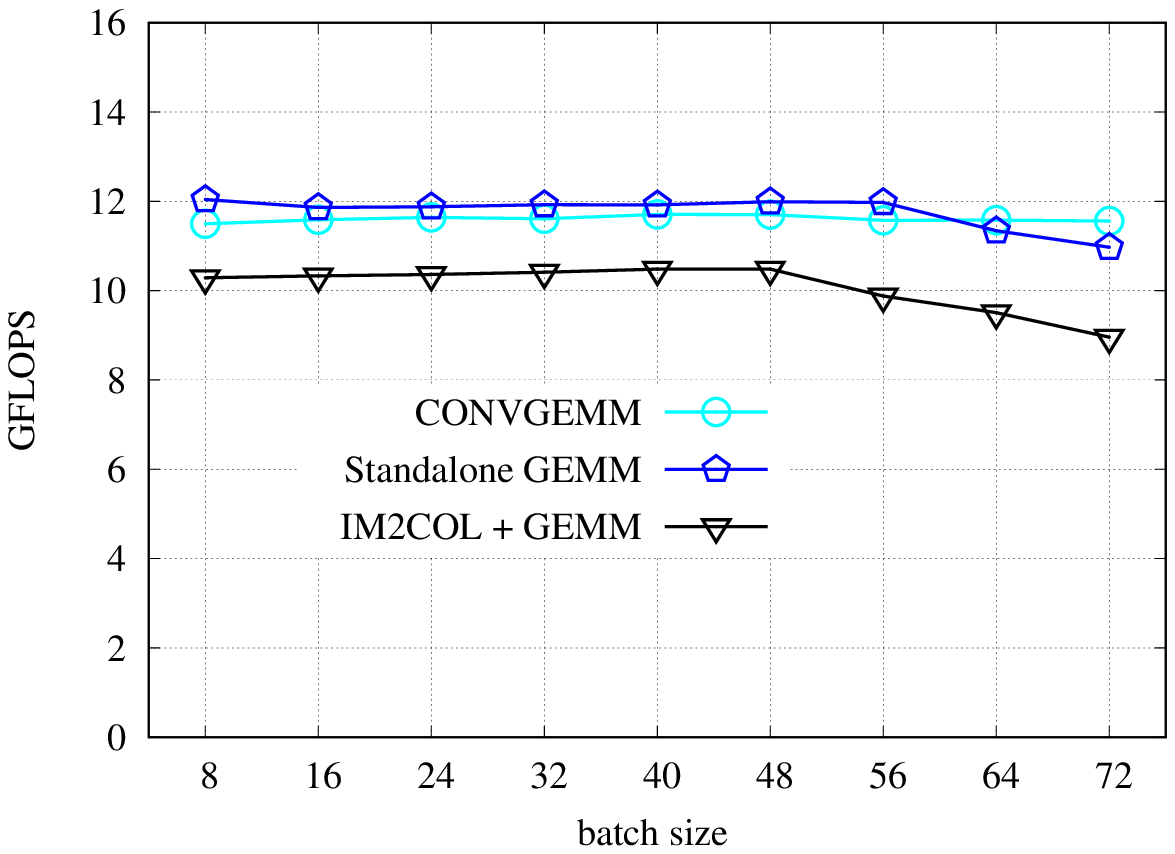}
  \caption{GFLOPS, VGG16, 1 core}
  \label{fig:results1tf}
  \end{subfigure} \\
\caption{Execution time (left column) and performance (right column) obtained by the indirect convolution algorithms for 
         AlexNet (top row), ResNet50 (middle row) and VGG16 (bottom row) on a single ARM Cortex-A57 core.}
\label{fig:results1t}
\end{figure}

\begin{figure}[p]
  \begin{subfigure}{0.5\textwidth}
  \centering
  \includegraphics[width=1.0\textwidth]{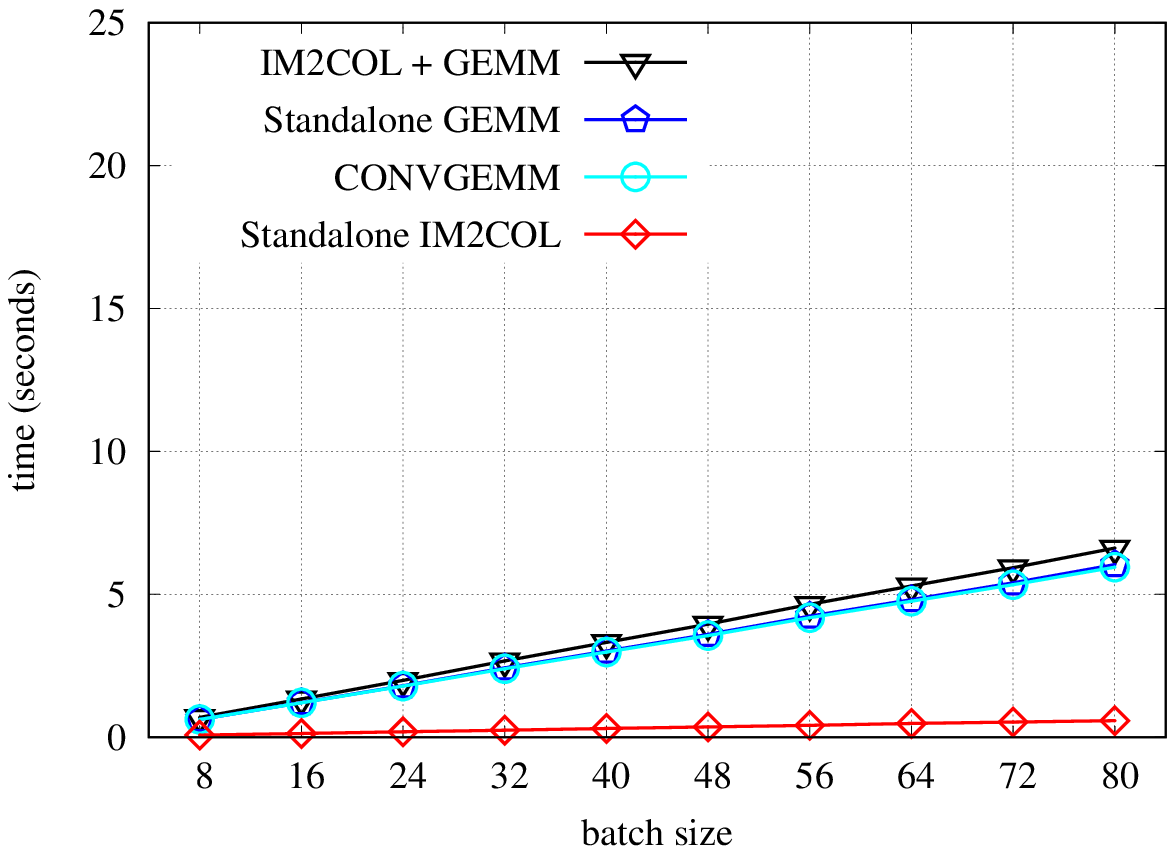}
  \caption{Execution time, AlexNet, 4 cores}
  \label{fig:results4ta}
  \end{subfigure}
  \begin{subfigure}{0.5\textwidth}
  \centering
  \includegraphics[width=1.0\textwidth]{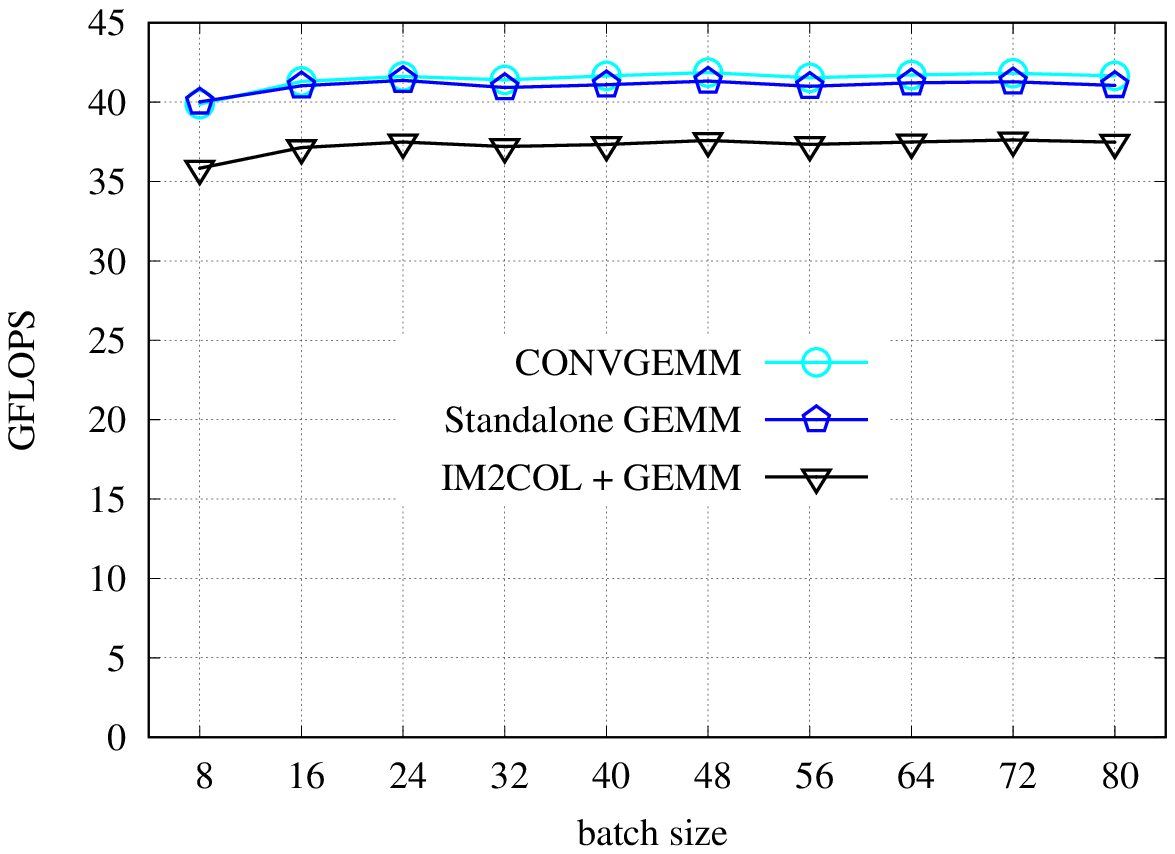}
  \caption{GFLOPS, AlexNet, 4 cores}
  \label{fig:results4tb}
  \end{subfigure} \\

  \begin{subfigure}{0.5\textwidth}
  \centering
  \includegraphics[width=1.0\textwidth]{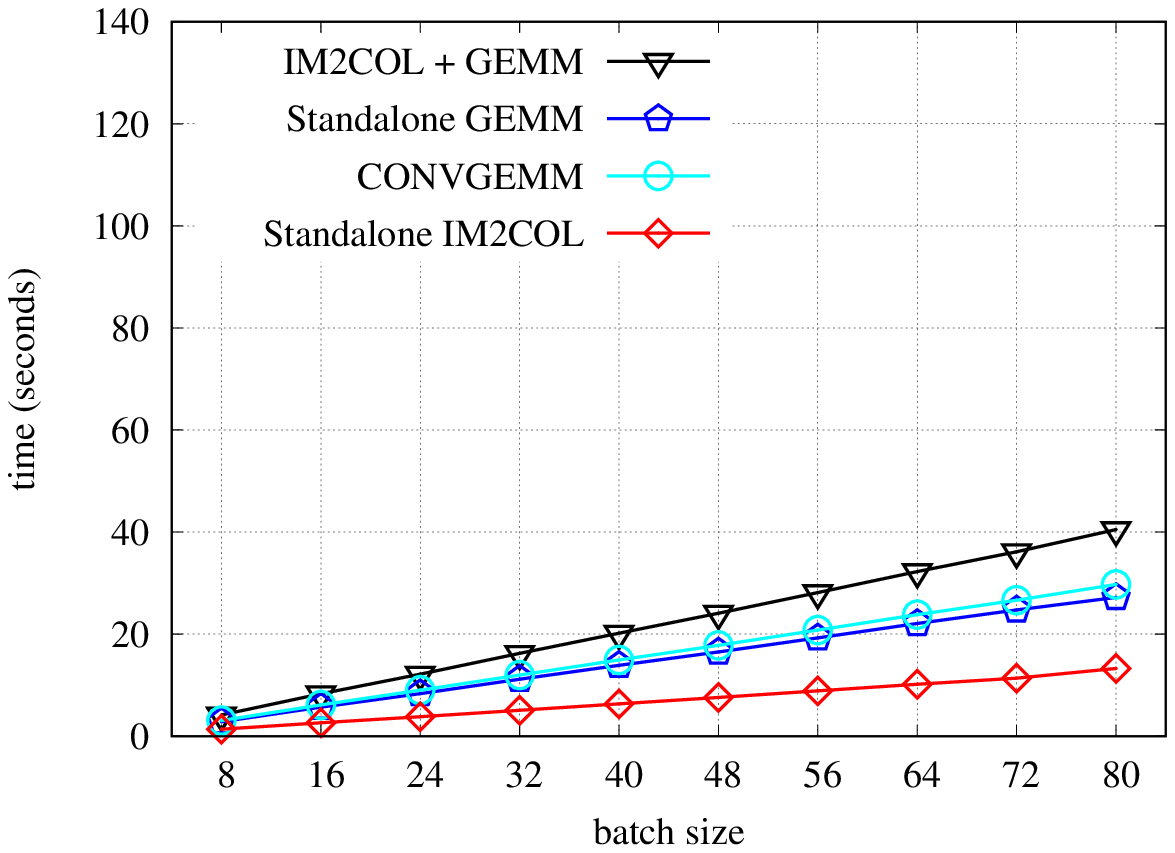}
  \caption{Execution time, ResNet50, 4 cores}
  \label{fig:results4tc}
  \end{subfigure}
  \begin{subfigure}{0.5\textwidth}
  \centering
  \includegraphics[width=1.0\textwidth]{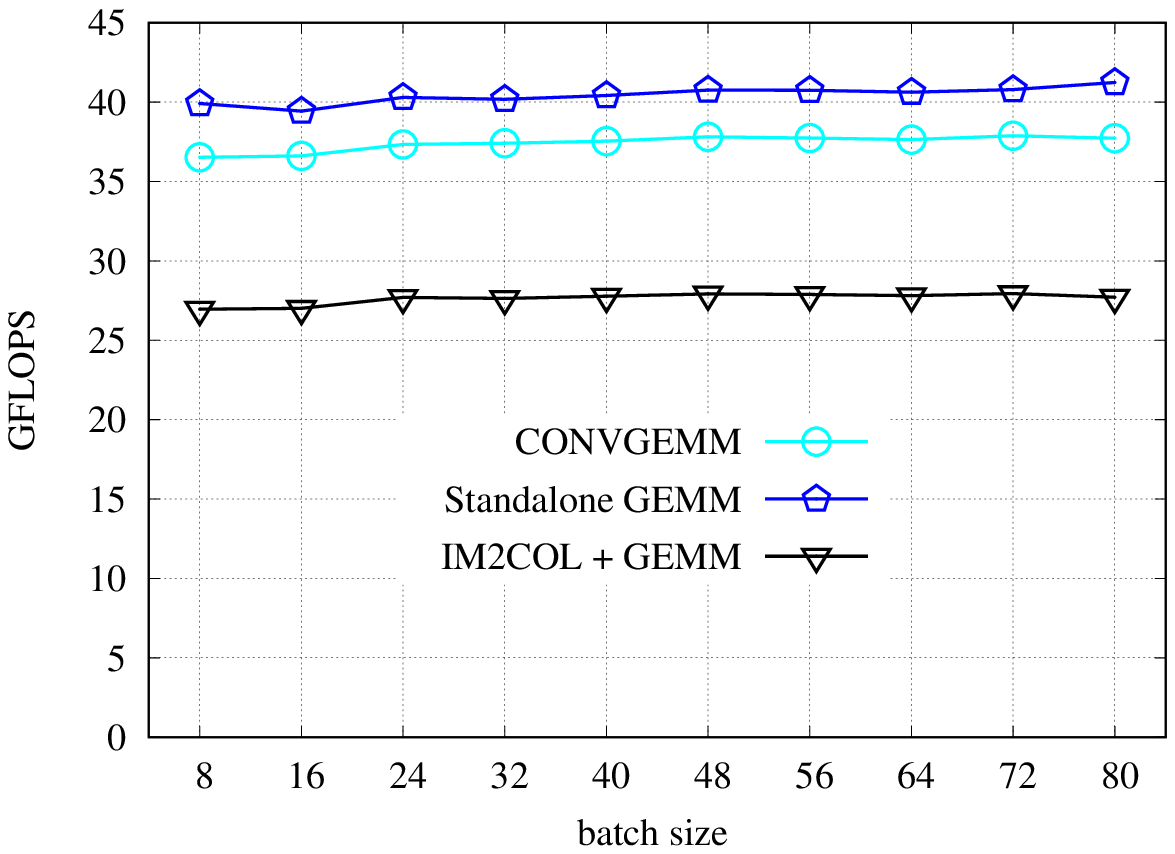}
  \caption{GFLOPS, ResNet50, 4 cores}
  \label{fig:results4td}
  \end{subfigure} \\

  \begin{subfigure}{0.5\textwidth}
  \centering
  \includegraphics[width=1.0\textwidth]{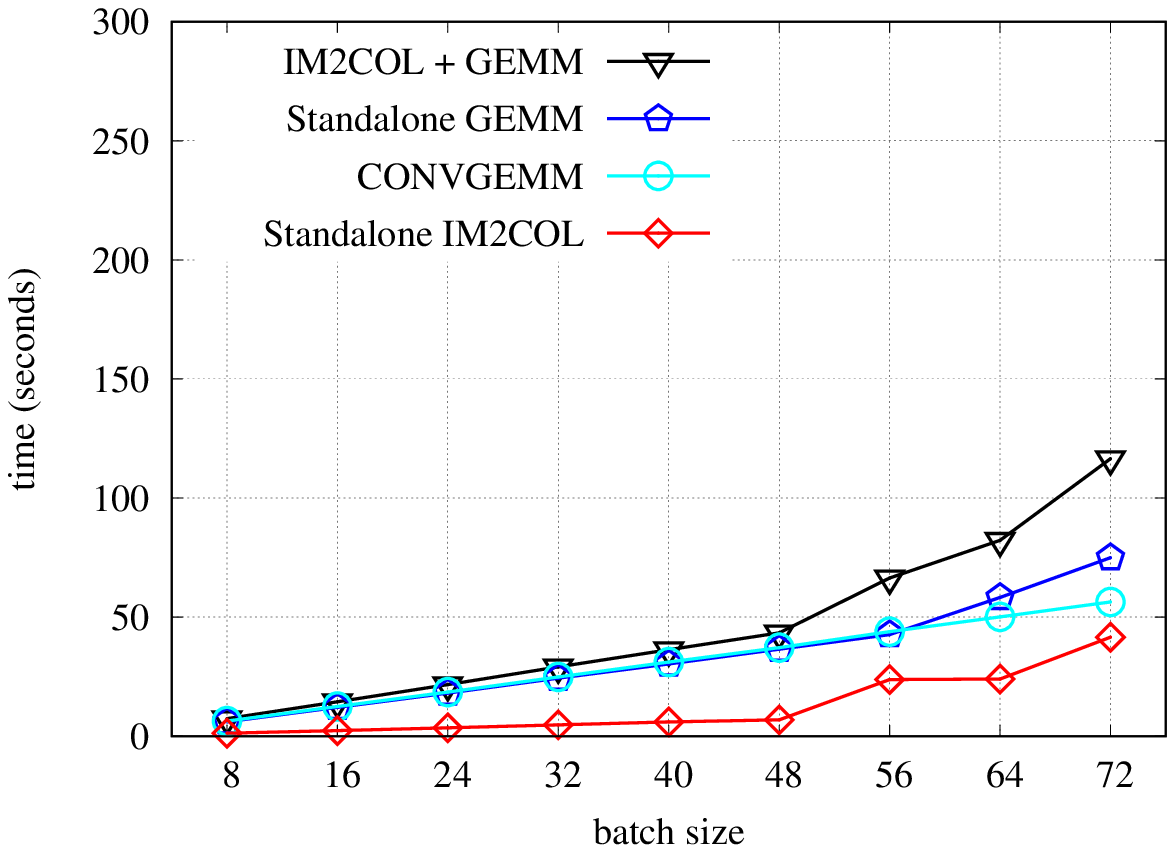}
  \caption{Execution time, VGG16, 4 cores}
  \label{fig:results4te}
  \end{subfigure}
  \begin{subfigure}{0.5\textwidth}
  \centering
  \includegraphics[width=1.0\textwidth]{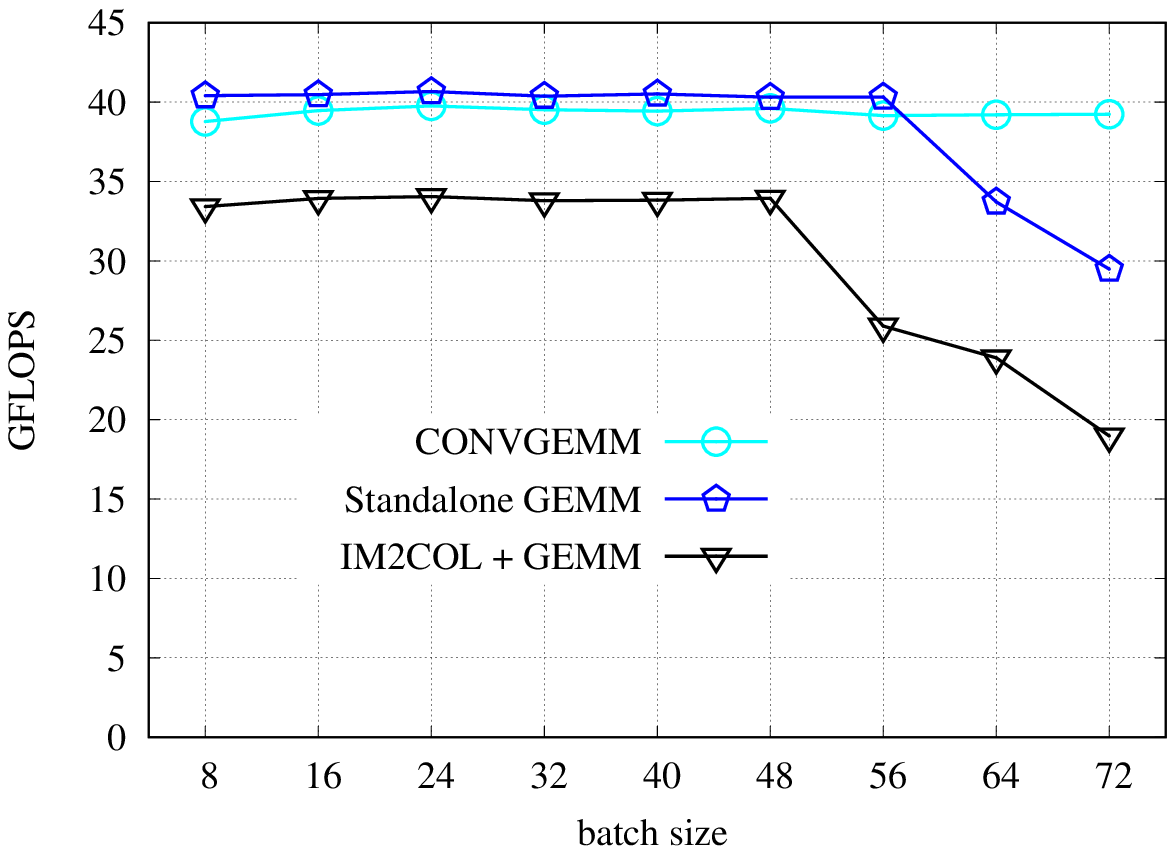}
  \caption{GFLOPS, VGG16, 4 cores}
  \label{fig:results4tf}
  \end{subfigure} \\
\caption{Execution time (left column) and performance (right column) obtained by the indirect convolution algorithms for 
         AlexNet (top row), ResNet50 (middle row) and VGG16 (bottom row) using all four ARM Cortex-A57 cores.}
\label{fig:results4t}
\end{figure}

Figures \ref{fig:results1t} and \ref{fig:results4t} show the time and performance (in GFLOPS, or billions
of floating-point operations per second) obtained for the evaluated models executed using a single core and the full 4-core processor, respectively. The plots display the execution time/performance attained for a range of batch sizes, for the optimized \convgemm algorithm  against the baseline approach (explicit) \imcoln+\gemm. 
In addition, all plots include the execution time/performance attained by the \gemm  kernels involved in the model simulation,
and the plots in the left-hand side include the time overhead required to perform the \imcol transforms.

For the AlexNet and ResNet50 models, the experiments are run up to a batch size $b=80$, while for VGG16 the largest value
for this parameter is only $b=72$. This is due to the large amount of memory required for the intermediate matrices assembled by 
the \imcol transform, which exceeds the memory capacity of the device (8 GiB) for the VGG16 model when $b=80$.

The results in  Figures~\ref{fig:results1t} and~\ref{fig:results4t} demonstrate that our technique
with an integrated \imcol fully hides cost of this transform for the AlexNet network, delivering the same execution time 
and GFLOPS rate observed when executing only the \gemm operations. When we tackle the two remaining 
(more complex) CNN models, the cost and performance of the optimized 
algorithm still remain close to those of the standalone \gemm operation while clearly 
outperforming the explicit \imcoln+\gemm counterpart.

There is a particular case worth of being discussed in some detail.
Concretely, 
for the explicit \imcoln+\gemm approach,
Figures~\ref{fig:results1tf} and~\ref{fig:results4tf} 
both show a notorious decrease in performance
for the VGG16 when $b>48$.
This decline is caused by the large size of the intermediate matrices, which results in I/O swapping to disk. 
The negative effect in the performance is more notorious in the multicore experiment, as in this case 
the memory access patterns performed during the explicit \imcol transform are more spread, increasing the effect of the swapping 
to disk. 

The observed negative effect in performance for large batch sizes and complex network models demonstrates that the optimized \convgemm algorithm, with an embedded \imcol, not only allows to perform the inference process for network models that cannot 
be tackled by the  explicit \imcoln+\gemm, but also avoids the efficiency pitfalls due to the earlier 
use of disk I/O in that approach.

\begin{figure}[!t]
  \begin{subfigure}{0.5\textwidth}
    \centering
    \includegraphics[width=\textwidth]{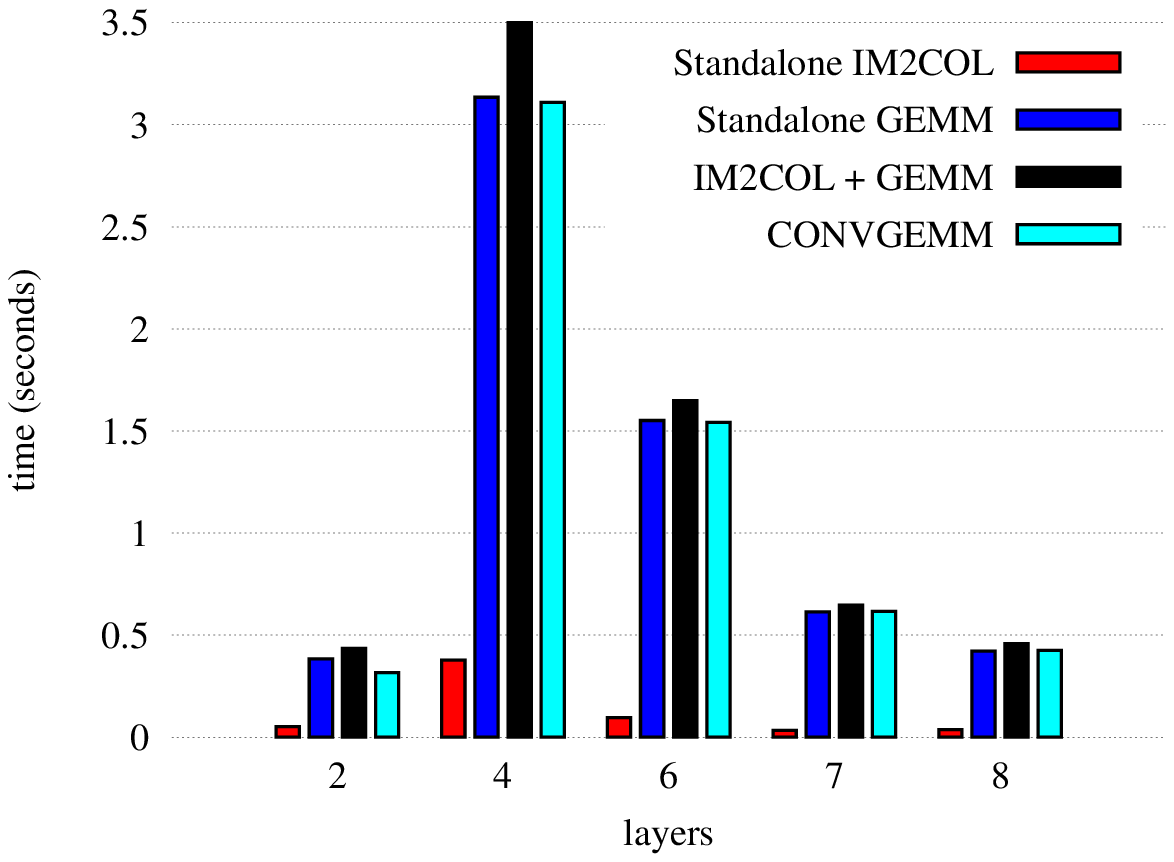}
  \end{subfigure}
  \begin{subfigure}{0.5\textwidth}
     \centering
      \includegraphics[width=\textwidth]{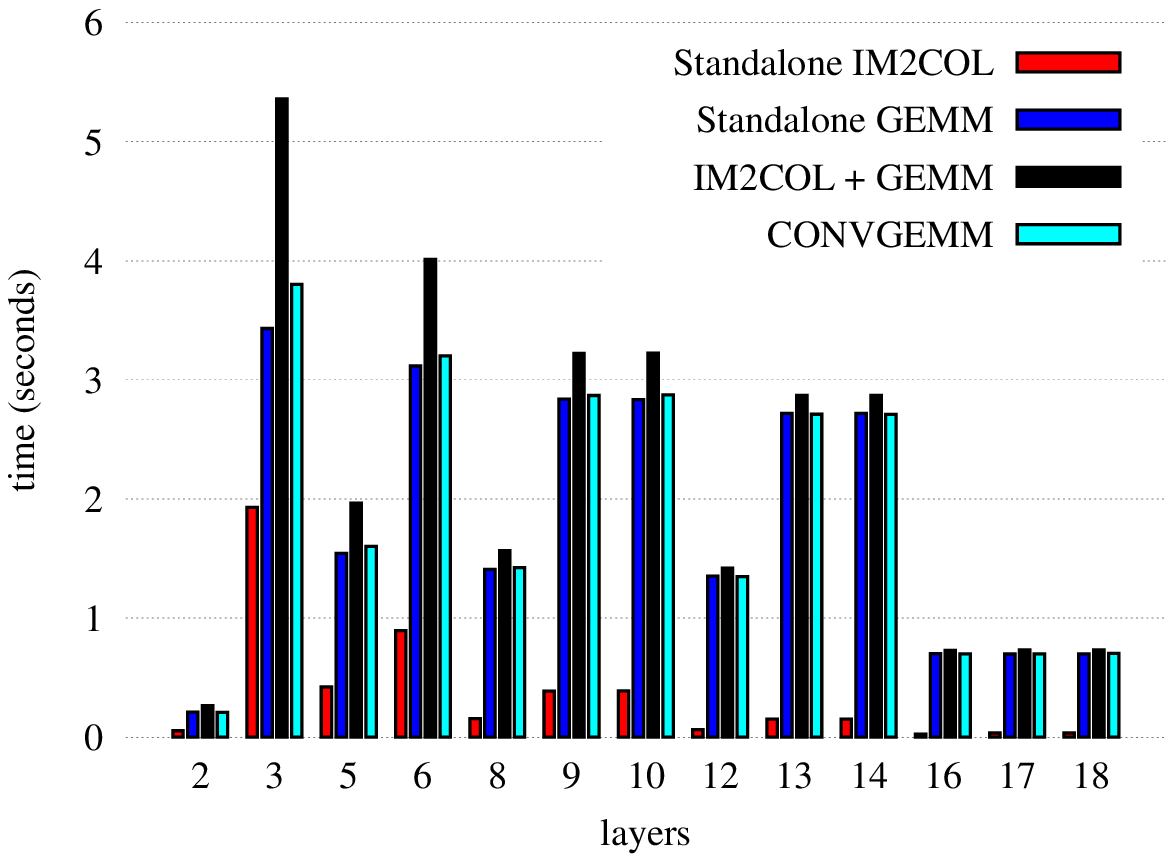}
  \end{subfigure}
  \caption{Execution time per layer obtained by the indirect convolution algorithms for AlexNet (left) and VGG16 (right) using all four ARM Cortex-A57 cores and a batch size $b=32$.}\label{fig:layerResults}
\end{figure}

To close the experimental analysis, Figure \ref{fig:layerResults} reports the execution time to compute the convolutions required at each CNN layer in the AlexNet and VGG16 models. The plots there illustrate that the time required per layer 
significantly varies between different layers.

\section{Closing Remarks}
\label{sec:remarks}

This work introduces a new convolution algorithm that outperforms the straight-forward \imcoln+\gemm approach in several
aspects. First, the new \convgemm algorithm removes the need of the additional memory work space 
utilized by the \imcoln+\gemm approach, enabling inference with large CNN models in memory bound systems.
In addition, the realization of the new scheme in combination with the BLIS kernel for \gemm yields an efficient
and portable implementation that can be migrated to other low-power architectures for which 
an optimized implementation of the BLIS micro-kernel exists (or can be developed).

The results in the experimental evaluation performed in this work show the remarkable performance advantage of the
new \convgemm scheme on a representative low-power ARM-based multicore processor, which completely eliminates the
workspace and performance overheads due to the utilization of an explicit \imcol transform.

\subsection*{Acknowledgements}
This research  was partially sponsored by projects 
TIN2017-82972-R of 
\textit{Ministerio de Ciencia, Innovaci\'on y Universidades}
and Prometeo/2019/109 of the
\textit{Generalitat Valenciana}.

\bibliographystyle{plain}
\bibliography{biblio,deep}

\begin{thebibliography}{10}

\bibitem{onednn}
{oneAPI} deep neural network library {(oneDNN)}: Performance library for deep
  learning, 2018.
\newblock Formerly known as Intel Math Kernel Library for Deep Neural Networks
  (Intel MKL-DNN) and Deep Neural Network Library (DNNL). Available from
  \url{https://oneapi-src.github.io/oneDNN/}.

\bibitem{cudnn}
Deep learning {SDK} documentation: {cuDNN} developer guide, 2020.
\newblock Available from
  \url{https://docs.nvidia.com/deeplearning/sdk/cudnn-developer-guide/index.html}.

\bibitem{mace}
Mobile {AI} {C}ompute {E}ngine documentation, 2020.
\newblock Available from \url{https://mace.readthedocs.io/en/latest/}.

\bibitem{nnpack}
{NNPACK}: Acceleration package for neural networks on multi-core {CPUs}, 2020.
\newblock Available from \url{https://github.com/Maratyszcza/NNPACK}.

\bibitem{DBLP:journals/corr/abs-1709-03395}
Andrew Anderson et~al.
\newblock Low-memory {GEMM}-based convolution algorithms for deep neural
  networks.
\newblock {\em CoRR}, abs/1709.03395, 2017.

\bibitem{DBLP:journals/corr/abs-1802-09941}
Tal Ben-Nun and Torsten Hoefler.
\newblock Demystifying parallel and distributed deep learning: An in-depth
  concurrency analysis.
\newblock {\em ACM Computing Surveys}, 52(4):65:1--65:43, August 2019.

\bibitem{Catalan2016}
Sandra Catal{\'a}n, Francisco~D. Igual, Rafael Mayo, Rafael
  Rodr{\'i}guez-S{\'a}nchez, and Enrique~S. Quintana-Ort{\'i}.
\newblock Architecture-aware configuration and scheduling of matrix
  multiplication on asymmetric multicore processors.
\newblock {\em Cluster Computing}, 19(3):1037--1051, 2016.

\bibitem{Che06}
Kumar Chellapilla, Sidd Puri, and Patrice Simard.
\newblock High performance {convolutional neural networks} for document
  processing.
\newblock In {\em International Workshop on Frontiers in Handwriting
  Recognition}, 2006.
\newblock Available as INRIA report inria-00112631 from
  \url{https://hal.inria.fr/inria-00112631}.

\bibitem{chetlur2014cudnn}
Sharan Chetlur, Cliff Woolley, Philippe Vandermersch, Jonathan Cohen, John
  Tran, Bryan Catanzaro, and Evan Shelhamer.
\newblock {cuDNN}: Efficient primitives for deep learning, 2014.
\newblock arXiv preprint 1410.0759. Available from
  \url{https://arxiv.org/abs/1410.0759}.

\bibitem{Cho17}
Minsik Cho and Daniel Brand.
\newblock {MEC}: Memory-efficient convolution for deep neural network.
\newblock In {\em Proceedings of 34th Int. Conference on Machine Learning --
  PMLR}, volume~70, pages 815--824, 2017.

\bibitem{recent-advances-in-deep-learning-for-speech-research-at-microsoft}
Li~Deng, Jinyu Li, Jui-Ting Huang, Kaisheng Yao, Dong Yu, Frank Seide, Mike
  Seltzer, Geoff Zweig, Xiaodong He, Jason Williams, Yifan Gong, and Alex
  Acero.
\newblock Recent advances in deep learning for speech research at {Microsoft}.
\newblock In {\em 2013 IEEE International Conference on Acoustics, Speech and
  Signal Processing}, pages 8604--8608, May 2013.

\bibitem{blas3}
Jack~J. Dongarra, Jeremy Du~Croz, Sven Hammarling, and Iain Duff.
\newblock A set of level 3 basic linear algebra subprograms.
\newblock {\em ACM Trans. on Mathematical Software}, 16(1):1--17, March 1990.

\bibitem{Dukh19}
Marat Dukhan.
\newblock The indirect convolution algorithm.
\newblock {\em CoRR}, abs/1907.02129, 2019.
\newblock arXiv preprint 1907.02129. Available from
  \url{https://arxiv.org/abs/1907.02129}.

\bibitem{qnnpack}
Marat Dukhan, Yiming Wu, and Hao Lu.
\newblock {QNNPACK}: open source library for optimized mobile deep learning,
  2020.
\newblock Available from \url{https://code.fb.com/ml-applications/qnnpack/}.

\bibitem{Geor18}
Evangelos Georganas, Sasikanth Avancha, Kunal Banerjee, Dhiraj Kalamkar, Greg
  Henry, Hans Pabst, and Alexander Heinecke.
\newblock Anatomy of high-performance deep learning convolutions on simd
  architectures.
\newblock In {\em Proceedings of the International Conference for High
  Performance Computing, Networking, Storage, and Analysis}, SC ’18, pages
  66:1--66:12. IEEE Press, 2018.

\bibitem{gitRepo}
{Source code repository}.
\newblock \url{https://gitlab.com/comtacts/convgemm}, 2020.

\bibitem{Goto:2008:HPI}
Kazushige Goto and Robert van~de Geijn.
\newblock High performance implementation of the level-3 {BLAS}.
\newblock {\em {ACM} Transactions on Mathematical Software}, 35(1):4:1--4:14,
  July 2008.

\bibitem{Goto:2008:AHP}
Kazushige Goto and Robert~A. van~de Geijn.
\newblock Anatomy of a high-performance matrix multiplication.
\newblock {\em {ACM} Trans. on Mathematical Software}, 34(3):12:1--12:25, May
  2008.

\bibitem{Han15}
Song Han, Huizi Mao, and William~J. Dally.
\newblock Deep compression: Compressing deep neural networks with pruning,
  trained quantization and {H}uffman coding, 2015.
\newblock arXiv preprint 1510.00149. Available from
  \url{https://arxiv.org/abs/1510.00149}.

\bibitem{he2016deep}
Kaiming He, Xiangyu Zhang, Shaoqing Ren, and Jian Sun.
\newblock Deep residual learning for image recognition.
\newblock In {\em Proceedings of the IEEE conference on computer vision and
  pattern recognition}, pages 770--778, 2016.

\bibitem{henryPacking}
Greg Henry.
\newblock {BLAS} based on block data structures.
\newblock Theory Center Technical Report CTC92TR89, Advanced Computing Research
  Institute. Cornell University, 1992.

\bibitem{jetson}
{NVIDIA Jetson TX2}.
\newblock
  \url{https://www.nvidia.com/es-es/autonomous-machines/embedded-systems/jetson-tx2/},
  2020.

\bibitem{Jin19}
Jintao {Ke}, Hai {Yang}, Hongyu {Zheng}, Xiqun {Chen}, Yitian {Jia}, Pinghua
  {Gong}, and Jieping {Ye}.
\newblock Hexagon-based convolutional neural network for supply-demand
  forecasting of ride-sourcing services.
\newblock {\em IEEE Trans. on Intelligent Transportation Systems},
  20(11):4160--4173, 2019.

\bibitem{Krizhevsky:2012:ICD:2999134.2999257}
Alex Krizhevsky, Ilya Sutskever, and Geoffrey~E. Hinton.
\newblock {ImageNet} classification with deep convolutional neural networks.
\newblock In {\em Proceedings of the 25th International Conference on Neural
  Information Processing Systems - Volume 1}, NIPS'12, pages 1097--1105, USA,
  2012. Curran Associates Inc.

\bibitem{Lavi16}
Andrew {Lavin} and Scott {Gray}.
\newblock Fast algorithms for convolutional neural networks.
\newblock In {\em 2016 IEEE Conference on Computer Vision and Pattern
  Recognition (CVPR)}, pages 4013--4021, 2016.

\bibitem{BLIS4}
Tze~Meng Low, Francisco~D. Igual, Tyler~M. Smith, and Enrique~S.
  Quintana-Ort\'{\i}.
\newblock Analytical modeling is enough for high-performance {BLIS}.
\newblock {\em ACM Trans. on Mathematical Software}, 43(2):12:1--12:18, August
  2016.

\bibitem{Ma18}
Ningning Ma, Xiangyu Zhang, Hai-Tao Zheng, and Jian Sun.
\newblock {ShuffleNet V2}: Practical guidelines for efficient {CNN}
  architecture design.
\newblock In {\em Proceedings European Conference on Computer Vision - ECCV
  2018. Lecture Notes in Computer Science}, volume 11218, pages 122--138, 2018.

\bibitem{Najafabadi2015}
Maryam~M. Najafabadi, Flavio Villanustre, Taghi~M. Khoshgoftaar, Naeem Seliya,
  Randall Wald, and Edin Muharemagic.
\newblock Deep learning applications and challenges in big data analytics.
\newblock {\em Journal of Big Data}, 2(1):1, Feb 2015.

\bibitem{OpenBLAS}
{OpenBLAS}.
\newblock \url{http://www.openblas.net}, 2015.

\bibitem{openmp08}
{OpenMP Architecture Review Board}.
\newblock {OpenMP} application program interface version 3.0, May 2008.

\bibitem{simonyan2014very}
Karen Simonyan and Andrew Zisserman.
\newblock Very deep convolutional networks for large-scale image recognition.
\newblock {\em arXiv preprint arXiv:1409.1556}, 2014.

\bibitem{BLIS3}
Tyler~M. Smith, Robert van~de Geijn, Mikhail Smelyanskiy, Jeff~R. Hammond, and
  Field~G. Van~Zee.
\newblock Anatomy of high-performance many-threaded matrix multiplication.
\newblock In {\em Proc. IEEE 28th Int. Parallel and Distributed Processing
  Symp.}, IPDPS'14, pages 1049--1059, 2014.

\bibitem{8114708}
Vivienne Sze, Yu-Hsin Chen, Tien-Ju Yang, and Joel~S. Emer.
\newblock Efficient processing of deep neural networks: A tutorial and survey.
\newblock {\em Proceedings of the IEEE}, 105(12):2295--2329, Dec 2017.

\bibitem{BLIS1}
Field~G. {Van~Zee} and Robert~A. {van~de~Geijn}.
\newblock {BLIS}: A framework for rapidly instantiating {BLAS} functionality.
\newblock {\em ACM Trans. on Mathematical Software}, 41(3):14:1--14:33, 2015.

\bibitem{ATLAS}
R.~Clint Whaley and Jack~J. Dongarra.
\newblock Automatically tuned linear algebra software.
\newblock In {\em Proceedings of the 1998 ACM/IEEE Conference on
  Supercomputing}, SC ’98, page 1–27, USA, 1998. IEEE Computer Society.

\bibitem{BLIS2}
Field G.~Van Zee, Tyler~M. Smith, Bryan Marker, Tze~Meng Low, Robert A. Van~De
  Geijn, Francisco~D. Igual, Mikhail Smelyanskiy, Xianyi Zhang, Michael
  Kistler, Vernon Austel, John~A. Gunnels, and Lee Killough.
\newblock The {BLIS} framework: Experiments in portability.
\newblock {\em ACM Trans. on Mathematical Software}, 42(2):12:1--12:19, June
  2016.

\bibitem{7243232}
Jiajun Zhang and Chengqing Zong.
\newblock Deep neural networks in machine translation: An overview.
\newblock {\em IEEE Intelligent Systems}, 30(5):16--25, Sep. 2015.

\bibitem{Zhan18}
Jiyuan Zhang, Franz Franchetti, and Tze~Meng Low.
\newblock High performance zero-memory overhead direct convolutions.
\newblock In {\em Proceedings of the 35th International Conference on Machine
  Learning -- PMLR}, volume~80, 2018.

\bibitem{Zhao18}
Yulin Zhao, Donghui Wang, Leiou Wang, and Peng Liu.
\newblock A faster algorithm for reducing the computational complexity of
  convolutional neural networks.
\newblock {\em Algorithms}, 11(10):159, Oct 2018.

\end{thebibliography}

\end{document}